\documentclass{aa}
\usepackage{epsfig}
\usepackage{graphicx}

\makeatletter
\def\gsim{\compoundrel>\over\sim}
\def\lsim{\compoundrel<\over\sim}
\def\compoundrel#1\over#2{\mathpalette\compoundreL{{#1}\over{#2}}}
\def\compoundreL#1#2{\compoundREL#1#2}
\def\compoundREL#1#2\over#3{\mathrel
      {\vcenter{\hbox{$\m@th\buildrel{#1#2}\over{#1#3}$}}}}
\makeatother

\begin{document}
   \title{Interstellar gas in the Galaxy and 
          the X-ray luminosity of Sgr ${\rm A}^{*}$ 
          in the recent past}

   \author{C. K. Cramphorn$^{1}$ 
           \and R. A. Sunyaev$^{1,2}$} 

   \offprints{conrad@mpa-garching.mpg.de}
 
   \institute{$^1$Max-Planck-Institut f\"ur Astrophysik,
              Karl-Schwarzschild-Str. 1, 
              85740 Garching bei M\"unchen, Germany\\
              $^2$Space Research Institute, Russian Academy of Sciences,
              Profsoyuznaya 84/32, 117810 Moscow, Russia\\}
   \date{Received ???; accepted ???}

   \abstract{Information about the X-ray 
    luminosity of the supermassive black hole located at the 
    Galactic center (GC), Sgr ${\rm A}^{*}$, 
    and its temporal variations in the past is imprinted 
    in the scattered emission observed today in the direction towards 
    giant molecular clouds (GMCs) located in our Galaxy. 
    Due to light travel time effects these clouds 
    probe the activity of Sgr ${\rm A}^{*}$ at different times 
    in the past depending on their position relative to the GC and the observer.
    In this paper we combine results of recent {\it ASCA} observations 
    along the Galactic plane, 
    providing upper limits for the
    scattered flux in the 4-10 keV range produced in a given direction, 
    with data from CO surveys of the 
    same regions. These CO surveys map the position and mass 
    of the molecular gas which the GMCs are made up of. 
    Demanding the scattered flux to be not larger
    than the observed one, this data enables us  
    to derive upper limits for the 4-10 keV luminosity of Sgr A$^{*}$  
    at certain times during the last $40\,000$ years 
    down to about $8\times10^{40}\,{\rm erg\,s}^{-1}$. 
    At other times the limits are less tight,
    of the order of $10^{41}-10^{42}\,{\rm erg\,s}^{-1}$.   
    For two periods of time of about $2000$ and 
    $4000$ years duration $8000$ and $14\,000$ years ago
    the currently available CO data is insensitive to any enhanced 
    activity of the GC. Flares lasting longer than 3000 years
    fill these time gaps and therefore can be excluded
    to have occured during the last $40\,000$ years with a luminosity
    larger than a few $10^{42}\,{\rm erg\,s}^{-1}$.
    The more extended and continuous HI distribution in the Galactic disk, which also 
    scatters the radiation emitted by Sgr A$^{*}$, 
    allows us to extend the time coverage 
    further into the past, back to about $110\,000$ years, 
    albeit the limits are becoming less tight.
    We thereby can rule out a long term X-ray activity phase of Sgr A$^{*}$ at
    one per cent of its Eddington level ending less than about $80\,000$ years ago.  
    The limits presented in this paper can be improved by  
    observations of emission in the fluorescent iron K$_{\alpha}$-line.
    We study the feasibility of these methods  
    to investigate past nuclear activity in other spiral galaxies 
    observed with the angular resolution of X-ray telescopes 
    like {\it Chandra} and {\it XMM-Newton}.
    }

   \titlerunning{The X-ray luminosity of Sgr ${\rm A}^{*}$ 
                 in the recent past}

   \maketitle

\section{Introduction}
High spatial resolution observations of the stars in the 
vicinity of the Galactic center (GC) strongly suggest the
presence of a supermassive black hole of mass
$2.6\times10^{6}\,M_\odot$ (Genzel et al.~1997; Ghez et al.~1998).
Recent {\it Chandra} measurements (Baganoff et al.~2001) 
detect the associated X-ray source Sgr ${\rm A}^{*}$
at a mere $2\times10^{33}\,{\rm erg}\,{\rm s}^{-1}$, 
more than ten orders of magnitude below the 
Eddington luminosity of 
$3\times10^{44}\,{\rm erg}\,{\rm s}^{-1}$ for a black hole
of this mass.
This relative dimness being remarkable in itself
raises the question whether Sgr ${\rm A}^{*}$
was equally dim in the past.
 
{\it Granat} spacecraft observations of the region surrounding the GC 
have revealed a close resemblance
between the morphology of the diffuse X-ray emission
above 10 keV
and the spatial distribution of molecular gas as
inferred from CO observations.  
Sunyaev et al.~(1993) proposed that the diffuse, hard 
X-ray emission observed today in the direction 
of giant molecular clouds (GMCs) in the central 100 pc of the Galaxy
and especially Sgr B2 is radiation emitted by Sgr A$^{*}$ in the past 
scattered into our line of sight by the molecular gas.
They predicted the presence of a strong
iron fluorescent K$_{\alpha}$-line at $6.4$ keV with an
equivalent width of the order of 1 keV. This line is a signature
of scattering of hard X-ray photons on neutral matter
when the primary photon source does not contribute to
the observed emission, by either falling outside of 
the observing beam or being currently not active 
(Vainshtein \& Sunyaev 1980).

In order for the mass of molecular gas 
($\sim6\times10^6\,M_{\odot}$)
present in Sgr B2 to 
produce the observed 
flux the X-ray luminosity of Sgr A$^{*}$ must have been 
of the order of 
$2 \times 10^{39} {\rm erg \, s}^{-1}$ about 400 years ago.
This time delay corresponds to the different path lengths for photons 
received directly or taking the detour via Sgr B2, which lies
at a projected distance of $\sim 120$ pc from the GC.

Evidence for the scattering scenario has been discovered 
with the spectroscopic imaging capabilities of the {\it ASCA} satellite
(Koyama et al.~1996; Sunyaev et al.~1998; 
Murakami et al.~2000).
The X-ray spectrum of Sgr B2 
exhibits strong emission
in the iron K$_{\alpha}$-line at $6.4$ keV confirming
the prediction of Sunyaev et al.~(1993).
Murakami et al.~(2001A) have recently observed
another GMC close to the GC, Sgr C, 
to be bright in the K$_{\alpha}$-line, strengthening
the case for the scattering scenario. Theses authors call 
Sgr B2 and Sgr C ``X-ray reflection nebulae''.
Recent {\it Chandra} observations (Murakami et al.~2001B) confirm
the {\it ASCA} results about Sgr B2 and the nature
of the 6.4 keV K$_{\alpha}$-line emission.

The application of the method successfully
used in the above described cases to constrain
the luminosity of Sgr A$^{*}$ in the past is not
restricted to GMCs close to the GC.
For it to provide nontrivial constraints 
upon the strength of the activity of the GC one 
needs a scattering material with a sufficient
optical depth and an estimate of the scattered 
flux produced by this material.
In the case of our Galaxy two obvious choices
which fulfill these criteria 
are interstellar molecular and atomic hydrogen:

\begin{list}{}{}
\item[a)]
CO surveys covering the Galactic plane have shown
that about $10^9\,M_{\odot}$ of H$_2$ reside in
GMCs on orbits between the GC and the sun
with a peak mass surface density around a galactocentric radius
between $4$ and $6$ kpc, the so called molecular ring (e.~g. Dame 1993).  
The time delays, 
the difference of the arrival times between the primary and 
the scattered radiation, 
for clouds located in the molecular ring
range up to about $40\,000$ years.

In this paper we will compute the scattered flux 
produced by a GMC in response to a strong flare of Sgr ${\rm A}^{*}$.
This flux is detectable by current X-ray telescopes even if it was  
relatively short, with a duration of the order of years.
Turning this argument around one infers from the observed relative 
dimness of GMCs today that there could not have been a strong flare 
of Sgr ${\rm A}^{*}$ at those times in the past, which correspond to
the time delays of massive GMCs in our Galaxy.

The good temporal ``resolution'' with an uncertainty of the occurrence 
of a flare of the order of years comes about because the mass of a GMC
is rather concentrated with a very dense core. We are therefore in 
principle able to pinpoint past flares of Sgr ${\rm A}^{*}$ 
during the last $40\,000$ years very well 
if there exists a GMC with a corresponding time delay.

The drawback of this ``accurate'' timing is that because there 
only exist a finite number of GMCs in our Galaxy it is
not possible to cover the whole time interval from 
$40\,000$ years ago, the time delay of the most distant 
clouds, until ``today''.
For the analysis presented in this paper it was only possible
to use the CO data from the first Galactic quadrant.
This means that we are missing all the GMCs, which are
located in the fourth Galactic quadrant and which should
be equally numerous. This additional data 
would probably improve the ``filling factor'' of our sample but 
still leave several time gaps and not extend the time coverage
beyond $40\,000$ years. 
The possibility thus always exists that a short 
flare of Sgr A$^{*}$ occured during the last $40\,000$ years 
unnoticed by us, because there is no GMC which could 
respond to this flare. 
If we take only the available CO data into account the 
coverage becomes complete for flares longer than about 
3000 years and we can thereby rule out any flare brighter 
than a few $10^{42}\,{\rm erg}\,{\rm s}^{-1}$ during 
the last $40\,000$ years of this or a longer duration. 

\item[b)]
The HI phase of the interstellar medium (ISM) is more 
diffusely distributed than H$_{2}$. It extends out to about 16 kpc with an
approximately constant mass surface density of about
$4\,M_\odot\,{\rm pc}^{-2}$, corresponding to a total mass of the order
of $3\times10^{9}\,M_\odot$ (e.~g. Dame 1993).
This implies that the total mass of neutral hydrogen in a field of view of the 
{\it ASCA} instrument is larger than the mass of any GMC even 
those near the GC. For a Galactic longitude of $10^{\mathrm{o}}$ it is
about $2\times 10^{7}\,M_{\odot}$.

Due to its more continuous distribution the response
of the HI distribution to flares of Sgr A$^{*}$ 
will be spread out over a long time interval.
Therefore we are unable to locate the time when a flare happened. 
A strong flare longer ago can produce the same response 
as a weak flare not so long ago.
Nevertheless due to the larger extent of HI in the Galactic disk 
we can extend the time coverage back to about $110\,000$ years.   
We will show that for example the switch off 
case from $1\%$ of the Eddington luminosity of
Sgr A$^{*}$
(the whole Galactic disk is illuminated by Sgr A$^{*}$,
which suddenly ``switches off'' at a time $t_\mathrm{o}$)  
can be ruled out to have ended less than about $80\,000$ years ago. 
\end{list}

The {\it ASCA} Galactic Plane Survey (Sugizaki et al.~2001),
provides upper limits for the diffuse X-ray emission produced 
by scattering along certain directions.  
In this paper we compute the expected scattered flux for an 
assumed Eddington luminosity of 
$3\times 10^{44} \,{\rm erg}\,{\rm s}^{-1}$ in the 4-10 keV
range. Although this value is rather large
the results can be rescaled to any input luminosity 
because the problem is linear.
It is obvious that the ratio of 
the computed scattered flux to
the upper limit for the scattered flux given by 
{\it ASCA} observations,  
$f=F_\mathrm{sc}/F_\mathrm{obs}$, gives
us an estimate of the average luminosity of the GC,
$L_\mathrm{GC}\lsim f^{-1} L_\mathrm{Edd}$, at a time which 
corresponds to the time delay of the observed GMC.
 
This paper is structured as follows. In the following section
we recall the basic ideas and relations of X-ray archaeology.
We describe how we extracted the CO and X-ray data 
necessary for our purposes from the literature.
In Sec.~3 we will derive constraints 
for the past X-ray luminosity of Sgr ${\rm A}^{*}$ 
using a sample of GMCs in the Galaxy.  
In Sec.~4 we will describe the limits the 
Galactic HI distribution can provide.
The possibility to use the same methods to study
past nuclear activity in other spiral galaxies is studied in Sec.~5. 
In Sec.~6 we close with a discussion of our results.

\section{Method}
\begin{figure}
\centering
\includegraphics[width=\columnwidth]{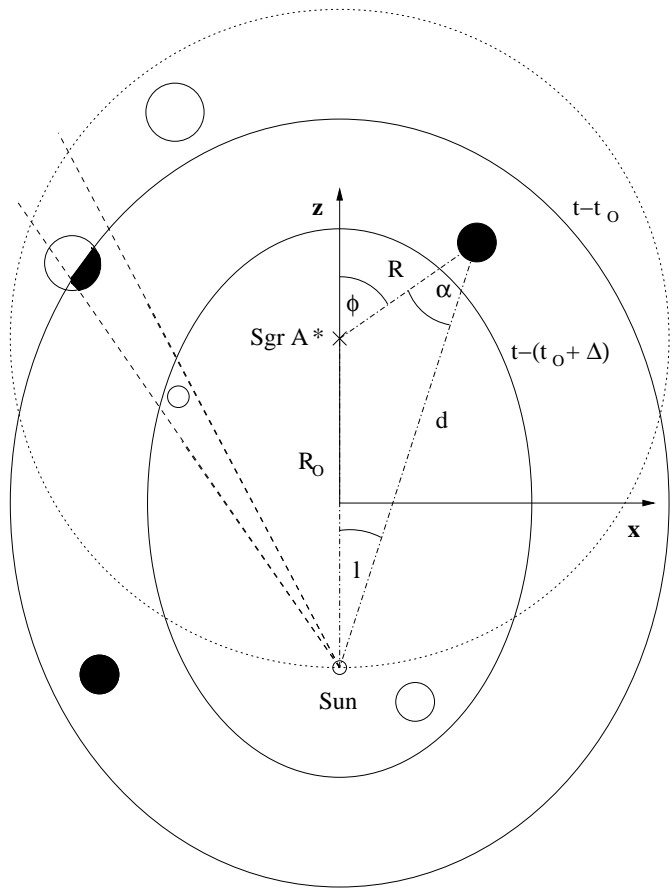}
 \caption{A sketch of the scenario studied  
          projected onto the Galactic plane. 
          A time $t_\mathrm{o}$ ago the luminosity of Sgr A$^{*}$ 
         (marked by a cross) flared. 
          This enhanced activity lasted for a period $\Delta$.
          Relative to the earth bound
          observer (small circle), observing at a time t and
          orbiting Sgr A$^{*}$ on the dotted circle, 
          the GMCs that lie between the two ellipsoids corresponding
          to the beginning and the end of the flare  
          are exposed to this enhanced emission.
          For other GMCs the period of enhanced 
          irradiation already has passed or has not started yet.
          The boundaries of an observational field of view are sketched
          by dashed lines. Inside this field of view one cloud has
          already faded and another one is partially covered by the 
          flare. Two possible photon paths between Sgr A$^{*}$ and
          the observer are marked by dot-dashed lines (one of them
          covered half way by the z-axis). 
          The ellipsoids are rotationally symmetric about the z-axis.}
\label{Fig1}
\end{figure}
\subsection{X-ray Archaeology}
The scenario we have in mind is sketched in Fig.~1.
An outburst of duration $\Delta$ has occured 
at the GC a time $t_\mathrm{o}$ ago, as measured by the earth 
bound observer. As time evolves this signal moves
radially outwards from the GC and starts illuminating GMCs 
at larger and larger galactocentric distances. These clouds
scatter a fraction of the incident radiation into the
direction of the observer.

The loci of scattering sites giving a fixed time delay form an 
ellipsoid with foci at the source and the observer. 
Since the system is symmetric around the line
connecting the GC and the observer 
this ellipsoid is fully described by two parameters 
which depend upon time past since the detection
of the burst $t-t_\mathrm{o}$, where $t$ is the time
of observation. In this section we will assume from now
on $t_\mathrm{o}=0$ to simplify the following formulae.

Choosing a cartesian coordinate system (see Fig.~1)
with the GC and the
Sun at $z=R_\mathrm{o}/2$ and $z=-R_\mathrm{o}/2$ respectively,
the ellipsoid can be expressed as 
\begin{equation}
\frac{z^2}{a^2}+\frac{x^2+y^2}{b^2}=1,
\end{equation}
where the lengths of the major and minor axis are 
$a=(ct+R_\mathrm{o})/2$ and
$b=\sqrt{(ct)^2+2ctR_\mathrm{o}}/2$,
with $c$ the speed of light and $R_\mathrm{o}$ the 
distance to the GC.

An alternative description of the ellipsoid
can be given in polar coordinates $R$, $\phi$ (defined in Fig.~1):
\begin{equation}
R(\phi)=\frac{p}{1+e\cos \phi},
\end{equation}
with $p=b^2/a$ and $e=R_\mathrm{o}/(2a)$.
Note that in the limit of the observer at infinity $(R_\mathrm{o}
\rightarrow \infty, p \rightarrow c t, e \rightarrow 1)$  
this ellipsoid becomes a paraboloid
\begin{equation}
R(\phi)=\frac{ct}{1+\cos \phi}.
\end{equation}
This approximation is valid for GMCs at the GC 
($R_\mathrm{o}\gg c t$)
and has been studied in the context of Sgr B2
by Sunyaev \& Churazov (1998).\footnote{The study of
light echoes has a long  history, starting with an explanation
of observations of Nova Persei 1901 by Couderc (1939). 
Since then it has been applied to several astrophysical 
objects, e.~g. 
supernova lightcurves (Morrison \& Sartori 1969),
scattering by interstellar dust (Alcock \& Hatchett 1978),
scattering of an 
isotropic (Sunyaev 1982) or anisotropic radio source (Gilfanov et al.~1987)
by the thermal gas in a cluster of galaxies,
reverberation mapping in AGN (e.~g. Peterson 1993) and
Compton echoes from gamma-ray bursts (Madau et al.~2000) to name a
few.} In cartesian coordinates the paraboloid is described by:
\begin{equation}
z=\frac{1}{2}
\left(
ct-\frac{x^2+y^2}{ct}
\right)+
\frac{R_\mathrm{o}}{2}.
\end{equation}

The velocities of the surfaces of constant time delay 
along the major and minor axes of the ellipsoid
are obtained by differentiation of the formulae for a and b.
The major axis 
increases at a constant speed of $c/2$. The velocity
along the minor axes depends upon time as follows  
\begin{equation}
\dot b=\frac{c}{2}\frac{ct
+R_\mathrm{o}}{\sqrt{(ct)^2+
2ctR_\mathrm{o}}}.
\end{equation}
At times small compared to $R_\mathrm{o}/c$ 
this velocity is larger than the velocity of 
light and approaches the value $c/2$ for $ct \gg R_\mathrm{o}$. 
Inside the Galaxy, where most of the GMCs lie inside the solar
circle ($R < R_\mathrm{o}$),  
this velocity is always larger than 
$(2/\sqrt{3})(c/2)\approx0.58 c$.

The velocity in a general direction $\phi$ is given by the
temporal derivative of Eq.~2 which yields
\begin{equation}
\dot R(\phi)=\frac{c}{2}\,
\frac{c^2 t^2 + 2 R_\mathrm{o}(c t +R_\mathrm{o})(1 +  \cos \phi)}
{(c t+R_\mathrm{o}(1+\cos \phi))^2}. 
\end{equation}
In the paraboloid approximation this becomes, 
by either differentiating Eq.~3 with respect to $t$ or
by taking the limit $R_\mathrm{o}\gg c t$ of Eq.~6,
\begin{equation}
\dot R(\phi)=\frac{c}{1+\cos\phi}.
\end{equation}
The apparent velocity is given by the projection 
onto the plane of the sky, which reads
\begin{equation}
v_\mathrm{app}=\frac{c\sin\phi}{1+\cos\phi}.
\end{equation}
This equation is equivalent to the formula for apparent
superluminal motion of matter having velocities close
to the speed of light, which is normally expressed in terms of the
angle $\theta=\pi-\phi$.
The formula for the apparent velocity in the Galactic plane
in the ellipsoid case is given by
\begin{equation}
v_\mathrm{app}=\frac{c}{2}\,
\frac{c^2 t^2 + 2 R_\mathrm{o}(c t +R_\mathrm{o})(1 +  \cos \phi)}
{(c t+R_\mathrm{o}(1+\cos \phi))^2}\,\sin (\phi-l). 
\end{equation}
One obtains an additional dependence upon $l$,
because the projection angle for a given direction $\phi$
depends upon galactocentric distance $R$ and therefore upon
time.
 
The average time delay of photons scattered by a cloud 
with galactocentric radius $R$ and distance 
from the observer $d$ is given by:
\begin{equation}
\delta=(R+d-R_\mathrm{o})/c.
\end{equation}
It is obvious that for clouds with a constant galactocentric distance 
$R$ those, which lie behind the GC, produce the largest
time delay $\delta_\mathrm{max}(R={\rm const.})=2R/c$. 
For clouds with a distance of 6 kpc  
to the GC, the edge of the molecular ring, the largest 
time delay is about $40\,000$ years. 
The more diffuse HI phase of the interstellar medium (ISM) 
extends up to radii of about 16 kpc (Dame 1993). 
Matter at these distances can in principle probe 
the activity of the GC about $110\,000$ years ago.

\subsection{GMCs in the Galactic disk}
\begin{figure}
\centering
\includegraphics[width=11cm]{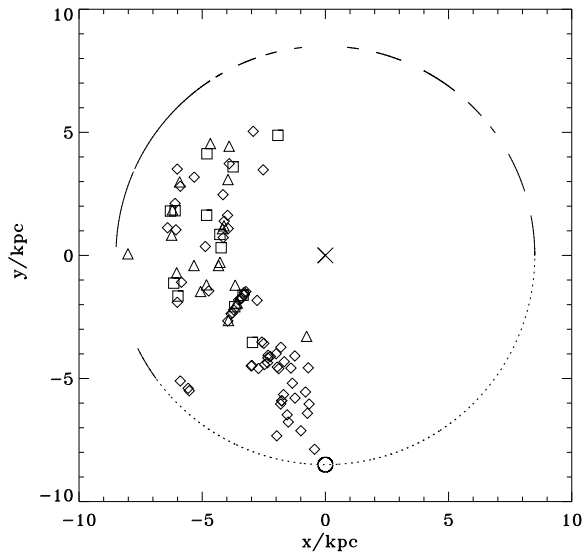}
 \caption{The spatial distribution of GMCs selected
          from SRBY87 for this work. 
          The GMCs cover a range of Galactic 
          longitudes of $8^{\mathrm{o}}\leq l 
          \leq 60^{\mathrm{o}}$.
          The position of the GC is marked by a
          cross. The position of the sun (marked by a small circle)
          is at coordinates $(-8.5,0.0)$. 
          For Galactic longitudes covered by {\it ASCA}
          observations the solar circle is marked by
          solid arcs. For Galactic longitudes falling outside
          of the {\it ASCA} survey the solar circle is marked 
          by a dotted line.
          The different symbols denote different cloud masses:
          $M_\mathrm{cl}\leq 5\times10^{5}\,M_\odot$ (diamonds),
          $5\times10^{5}\,M_\odot<M_\mathrm{cl}\leq10^{6}\,M_\odot$ (triangles)
          and $M_\mathrm{cl} > 10^{6}\,M_\odot$ (squares).}
\label{Fig2}
\end{figure}

The results of several CO surveys have been published over the last decades. 
They map the CO brightness temperature as a function of Galactic
longitude $l$,
Galactic latitude $b$ and radial velocity component $v_{r}$, 
differing in resolution and area surveyed.

Solomon et al.~(1987, SRBY87) have published a 
list (see Table 1 in their paper) of 273 molecular clouds 
detected with the Massachusetts-Stony Brook CO Galactic Plane Survey
of the first Galactic quadrant.
These clouds were extracted from the three dimensional 
$(l,b,v_{r})$ data cube by locating surfaces of
equal brightness temperature.
The distances were computed from the radial 
velocity component of the centroids under the usual 
assumption (see e.g. Binney \& Merrifield 1998) that clouds 
move on circular orbits with a constant circular velocity
of $\Theta_\mathrm{o}=220\, {\rm km\,s}^{-1}$.

The assumption of circular orbits with 
a constant rotational velocity is a fair approximation
for clouds inside of the solar circle but breaks down at small
galactocentric radii because the orbits are probably
highly non circular in this region.
For GMCs inside the solar circle a ambiguity is
inherent to this method of distance determination. 
A cloud with a given measured radial velocity component
can either lie on the near or the far side of the circular 
orbit it is assumed to travel around the GC. 
SRBY87 were able to overcome this ambiguity by several
methods described in their paper.

\begin{table*}
\centering
\begin{tabular}{c c c c c c c c c c c c} 
GMC & $l$/deg & $b$/deg & $R$/kpc & $d$/kpc & $\delta^{\mathrm{a}}$&
$M_\mathrm{cl}^{\mathrm{b}}$ & $r_\mathrm{cl}$/pc &
$t_\mathrm{cl}^{\mathrm{a}}$&
$F_{\mathrm{sc}}^{\mathrm{c}}$ & 
$F_{\mathrm{obs}}^{\mathrm{d}}$ &
$L_\mathrm{GC}^{\mathrm{e}}$\\
\\
003 & 08.30 & +0.00 & 3.4 & 5.3 & 669 & 55.6 & 19.3 &44 
&$7.1\times10^{-8}$&$1.9\times10^{-11}$&$8.0\times10^{40}$\\ 
014 & 10.00 & -0.04 & 4.6 & 4.0 & 335 & 11.3 & 5.8  &12
&$1.2\times10^{-8}$&$1.5\times10^{-11}$&$3.8\times10^{41}$\\ 
059 & 17.20 & -0.20 & 5.2 & 12.7 & 31449 & 199.1 & 29.7 &353 
&$9.4\times10^{-9}$&$1.7\times10^{-11}$&$5.4\times10^{41}$\\
080 & 20.75 & +0.10 & 4.2 & 5.1 & 2677 & 21.6 & 8.9 &39
&$1.0\times10^{-8}$&$1.5\times10^{-11}$&$4.5\times10^{41}$\\
085 & 21.75 & +0.00 & 4.6 & 4.5 & 2007 & 36.4 & 9.4 & 37
&$1.6\times10^{-8}$&$1.6\times10^{-11}$&$3.0\times10^{41}$\\
089 & 22.05 & +0.20 & 5.4 & 3.6 & 1673 & 33.7 & 11.9 &38
&$2.5\times10^{-8}$&$1.6\times10^{-11}$&$1.9\times10^{41}$\\
116 & 24.60 & -0.15 & 4.4 & 10.3 & 20743 & 106.7 & 13.4 &126 
&$5.2\times10^{-9}$&$1.5\times10^{-11}$&$8.7\times10^{41}$\\
122 & 25.65 & -0.10 & 4.2 & 9.8 & 18401 & 202.6 & 28.2 &283
&$1.7\times10^{-8}$&$1.8\times10^{-11}$&$3.2\times10^{41}$\\
128 & 25.90 & -0.15 & 3.8 & 7.7 & 10037 & 107.5 & 17.2 &133
&$1.2\times10^{-8}$&$1.7\times10^{-11}$&$4.3\times10^{41}$\\
151 & 28.30 & -0.10 & 4.8 & 4.9 & 4015 & 41.7 & 12.5 &54
&$1.3\times10^{-8}$&$1.4\times10^{-11}$&$3.2\times10^{41}$\\
152 & 28.60 & +0.05 & 4.3 & 7.5 & 11041 & 63.3 & 11.7 &90
&$5.4\times10^{-9}$&$1.4\times10^{-11}$&$7.8\times10^{41}$\\
158 & 29.00 & +0.05 & 4.3 & 7.5 & 11041 & 48.9 & 18.0 &145
&$6.4\times10^{-9}$&$1.5\times10^{-11}$&$7.0\times10^{41}$\\
162 & 29.85 & -0.05 & 4.3 & 7.4 & 10706 & 85.1 & 15.8 &127
&$8.5\times10^{-9}$&$1.5\times10^{-11}$&$5.3\times10^{41}$\\
171 & 30.80 & -0.05 & 4.6 & 5.8 & 6357 & 119.3 & 22.1 &107
&$2.3\times10^{-8}$&$1.3\times10^{-11}$&$1.7\times10^{41}$\\
191 & 33.90 & +0.10 & 4.8 & 7.1 & 11375 & 54.8 & 15.6 &119
&$6.0\times10^{-9}$&$1.0\times10^{-11}$&$5.0\times10^{41}$\\
193 & 34.25 & +0.10 & 6.1 & 3.1 & 2342 & 39.4 & 9.3 &42
&$1.6\times10^{-8}$&$1.0\times10^{-11}$&$1.9\times10^{41}$\\
201 & 35.20 & -0.10 & 7.9 & 0.8 & 669 & 2.0 & 3.4 &18
&$1.6\times10^{-8}$&$9.1\times10^{-12}$&$1.7\times10^{41}$\\
206 & 36.40 & -0.10 & 6.3 & 3.1 & 3011 & 23.2 & 8.8 &42
&$1.1\times10^{-8}$&$1.0\times10^{-11}$&$2.7\times10^{41}$\\
214 & 39.85 & -0.20 & 6.3 & 9.6 & 24758 & 375.3 & 47.4 &469
&$1.6\times10^{-8}$&$7.7\times10^{-12}$&$1.4\times10^{41}$\\
217 & 41.15 & -0.20 & 6.2 & 9.1 & 22750 & 201.2 & 22.4 &215
&$6.7\times10^{-9}$&$8.6\times10^{-12}$&$3.9\times10^{41}$\\
\end{tabular}
\begin{list}{}{}
  \item[$^{\mathrm{a}}$] Time delay $\delta$ and time for the
                         ellipsoid to cross the cloud
                         $t_\mathrm{cl}$
                         in units of years
  \item[$^{\mathrm{b}}$] Mass of cloud in units of $10^4\,M_\odot$
  \item[$^{\mathrm{c}}$] Scattered flux in the 4-10 keV range in units
                         of ${\rm erg}\,{\rm cm}^{-2}\,{\rm s}^{-1}$
                         for the whole cloud covered by a flare 
  \item[$^{\mathrm{d}}$] Flux in the 4-10 keV range observed by {\it
                         ASCA} in units
                         of ${\rm erg}\,{\rm cm}^{-2}\,{\rm s}^{-1}$
  \item[$^{\mathrm{e}}$] Upper limit for the luminosity of Sgr
                         A$^{*}$ in the 4-10 keV range in units of ${\rm erg}\,{\rm s}^{-1}$
                         a time $\delta$ ago
\end{list}
\caption{Selected parameters (number of cloud in SRBY87 sample, Galactic
longitude, Galactic latitude, galactocentric distance, distance, 
time delay, mass, radius, time for ellipsoid to cross the cloud,
scattered flux, observed flux and 
upper limit for the luminosity of Sgr A$^{*}$) 
of twenty GMCs from our sample.}
\end{table*}

This data set is not the most recent one (for more recent surveys
of this area see, e.g. Dame et al.~(2001) and Lee et al.~(2001))
but to our knowledge up to today the only one where mass 
estimates of a large number of individual GMCs have been presented.

Not all of the clouds listed in SRBY87
fall into one of the fields of view of the {\it ASCA} 
Galactic Plane Survey. These are clouds located at longitudes
larger than $60^{\mathrm{o}}$ or at latitudes larger
than $0.3^{\mathrm{o}}$. 
Combining the two data sets leaves one with 96 clouds 
out of 273. The spatial distribution of these clouds 
in the Galactic plane is plotted in Fig.~2, symbolised according
to their mass (for the distribution of the full set of clouds
see Solomon \& Rivolo 1989). Although spiral structure might be
marginally discernible in this picture, the concentration of GMCs along
the molecular ring is clearly visible.

The cloud properties needed for this work are the 
Galactic longitude $l$, the galactocentric radius $R$, the distance 
from the observer $d$,
the radius $r_\mathrm{cl}$ and the mass $M_\mathrm{cl}$ of a cloud,
which was computed with the observed velocity dispersion
and should thus not be affected by recalibrations of
the I(CO)/N(H$_{2}$) ratio. 
For the subsample of clouds we are working with
these parameters fall in the ranges $1.9\times10^4\,M_\odot \lsim M_\mathrm{cl}
\lsim 3.8 \times 10^6\,M_\odot$, $2.5 \lsim r_\mathrm{cl} \lsim 47$ pc
and $3\times10^2 {\rm years} \lsim \delta \lsim 4\times 10^4 {\rm years}$.
Note that in SRBY87 a distance of the
sun to the GC of 10 kpc was assumed.
We have rescaled the distances and derived cloud properties
to $R_\mathrm{o}=8.5\,{\rm kpc}$. 
In Table 1 (columns 1-8) we list the important 
parameters for twenty clouds 
out of our sample. Note that the GMCs in SRBY87 have been ordered 
and numbered according to their Galactic longitude.

\subsection{Scattered flux}
The {\it ASCA} observations of the Galactic plane $(b\approx 0^{\mathrm{o}})$ 
in the longitude range $-45^{\mathrm{o}}\leq l \leq 63^{\mathrm{o}}$
have been compiled and analysed by Sugizaki et al.~(2001).
In Fig.~2 we have tried to illustrate the range of 
longitudes covered by the
{\it ASCA} Galactic Plane Survey by plotting solid arcs
at a distance $R_\mathrm{o}$ from the GC, where observations
have been made. It is obvious
that we are missing the CO data from the fourth Galactic quadrant.
As mentioned before surveys of this area exist (e.~g. Dame et al.~2001)
but to our knowledge this data has not yet been fully analysed.
The distribution of GMCs should be symmetric between the
Galactic quadrants I and IV. We should therefore expect
a similar number of GMCs in the fourth Galactic quadrant.

The {\it ASCA} observations have been presented in three energy bands.
Most interesting for this work because least affected by
photoelectric absorption is the X-ray surface brightness in the hardest, 
the 4-10 keV, spectral band.
The corresponding fluxes at all longitudes are all of a 
similar order of magnitude. Not even one cloud seems to 
be especially bright ``at the moment''.
 
Even if the diffuse X-ray flux observed in one field of view
originates from a different physical process than scattering,
it is still an upper limit for the flux produced through 
scattering by gas distributed
along this beam: $F_\mathrm{sc}\leq F_\mathrm{obs}$.

The clouds in our sample have optical depths
for Thomson scattering
up to $\tau_\mathrm{T}\approx10^{-1}$.
We therefore consider scattering in the optically thin limit.
Our treatment of absorption is described
in Sec.~2.~3.~1. Nearly the whole Galaxy is transparent to X-rays with energies
above 4 keV, except some regions around the GC. But even here
optically thick GMCs do not shield the whole disk of the Galaxy
from the illumination by Sgr A$^{*}$. The covering fraction is 
probably rather small.

In the optically thin case the scattered surface 
brightness in a given direction
at energy $E$ and time of observation $t$ is given by
\begin{equation}
I_\mathrm{sc}(E,t)=
\int_{s_\mathrm{min}(t)}^{s_\mathrm{max}(t)}
n(s)\frac{L_\mathrm{GC}(E,t_\mathrm{ret})}
{4\pi R(s)^2}
\left(\frac{{\rm d}\sigma}{{\rm d}\Omega}\right){\rm d}s, 
\end{equation}
where $s_\mathrm{min}$ and $s_\mathrm{max}$ are the boundaries of integration,
$n(s)$ is the number density of the scattering gas at a distance $s$ from
the observer,
$t_\mathrm{ret}=t-(R(s)+s)/c$ the ``retarded'' time, $R$ the distance to the GC  
and $({\rm d}\sigma/{\rm d}\Omega)$ the differential
scattering cross-section. The time dependent integration limits 
are the points of
intersection of the line of sight with the ellipsoids corresponding
to the beginning and the end of a flare.
 
The scattered flux is given by the integration of the surface
brightness over the field of view of the {\it ASCA} instrument.
We take into account the cases when parts of a cloud fall outside 
the field of view, as sketched in Fig.~1, 
or when the field of view is larger than a cloud:
\begin{equation}
F_\mathrm{sc}(E,t)=\int I_\mathrm{sc}(E,t){\rm d}\Omega.
\end{equation}
We neglect here the iron K$_\alpha$-line production
and the corresponding iron K-edge absorption.
The detailed {\it ASCA} spectral data are not yet publicly  
available. {\it XMM-Newton} will provide much improved
data including K$_\alpha$-line limits, which will improve
the limits for the X-ray luminosity of Sgr A$^{*}$
derived in the following sections.

In the optical thin case the scattered luminosity
is proportional to the mass of the illuminated scattering material.
An order of magnitude estimate of the above integrals
to demonstrate the scaling yields
\begin{equation}
L_\mathrm{sc}\sim
\frac{3}{16\pi}\frac{\sigma_\mathrm{T}}{R^2}
\frac{M_\mathrm{cl}}{m_\mathrm{p}}
L_\mathrm{GC}.
\end{equation}
This formula neglects the spatial extension of the scattering
gas and the angular dependence of the differential scattering
cross-section and assumes that the whole cloud is covered
by the flare, implying that all of the mass contributes
to the scattered emission.
For typical cloud parameters one obtains a
scattered luminosity of
\begin{equation}
L_{\mathrm{sc}}\sim
6\times10^{37}\,{\rm erg \, s}^{-1}
\left(\frac{L_\mathrm{GC}}{L_\mathrm{Edd}}
\right)
\left(\frac{M_\mathrm{cl}}{10^6\,M_{\odot}}
\right)
\left(\frac{5\,{\rm kpc}}{R}
\right)^2,
\end{equation}
which for $L_\mathrm{GC}\sim L_\mathrm{Edd}$ 
is comparable to the luminosity of the 
brightest X-ray binaries in the Galaxy, thereby
certainly contradicting the observations.

If the flare duration $\Delta$ is shorter than the 
time for the ellipsoid to sweep across the cloud
one has to multiply Eq.~14 by an additional factor
\begin{equation}
\left(\frac{\Delta}{2r_\mathrm{cl}/v_\mathrm{ell}}
\right),
\end{equation}
where $v_\mathrm{ell}$ is the velocity of
the ellipsoid across the cloud, which can be superluminal
depending on the relative orientation of the cloud with
respect to the GC and the observer. 
This additional factor takes into account that only a
fraction $\sim \Delta/(2r_\mathrm{cl}/v_\mathrm{ell})$ 
of the mass of the cloud is illuminated.

\subsubsection{Absorption}
As X-ray photons of energy $E$ traverse the ISM 
a fraction $\tau_\mathrm{ph}(E)$ of them is  
absorbed by the heavy elements of the interstellar gas,
again assuming optically thin conditions. 
Therefore a GMC will be exposed to a flux
$(L_\mathrm{gc}(E,t_\mathrm{ret})/4\pi R^2)
\exp[-\tau_\mathrm{ph}'(E)]$,
where $\tau_\mathrm{ph}'(E)$ is the optical
depth for photoelectric absorption at energy $E$ 
between the GC and the GMC.
The observer will measure a flux
$F_\mathrm{sc}(E,t)
\exp[-\tau_\mathrm{ph}(E)]$.
The optical depth due to photoelectric absorption is proportional
to the hydrogen column density $N_\mathrm{H}$
\begin{equation}
\tau_\mathrm{ph}(E)=N_\mathrm{H}\sigma_\mathrm{ph}(E),
\end{equation}
where $\sigma_\mathrm{ph}(E)$ is the photoelectric absorption 
cross-section per hydrogen atom, which we compute
with an analytic fit by Morrison \& McCammon (1983).

We estimate $N_\mathrm{H}$ with the average 
mass surface density of a GMC given by 
$\Sigma_\mathrm{cl}\approx M_\mathrm{cl}/r_\mathrm{cl}^2\pi$,
which leads to
$N_\mathrm{H}\approx (3/4)(\Sigma_\mathrm{cl}/m_\mathrm{p})$.
Since we have no better knowledge about the hydrogen column
density between the GC and a cloud, 
we furthermore assume $N_\mathrm{H}'=N_\mathrm{H}$.

\subsubsection{Differential scattering cross-section for a mixture of H$_2$ and He}
\begin{figure}
\centering
\includegraphics[width=\columnwidth]{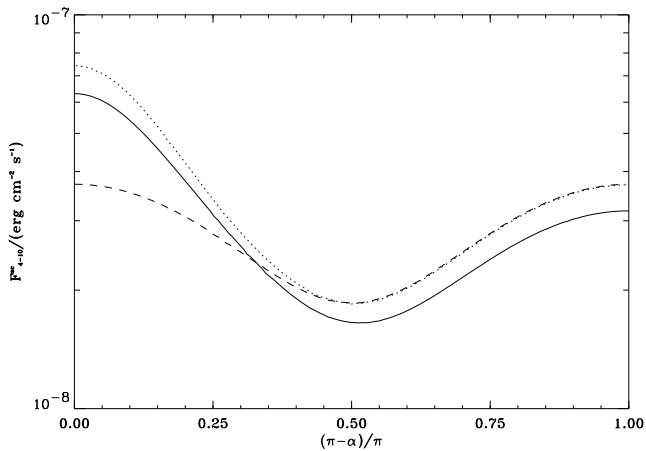}
 \caption{The scattered flux in the 4-10 keV 
          as a function of the angle $\alpha$
          (defined in Fig.~1) for a cloud of $10^6\,M_\odot$
          at 5 kpc distance from the GC and the observer
          fully illuminated by an Eddington flare of Sgr A$^{*}$.
          Note that $\alpha=\pi$ corresponds to
           the case of forward scattering.
          The lines mark different mass compositions
          of the cloud: 
          70\% H$_{2}$ + 30\% He (solid line),  
          atomic hydrogen (dashed) and
          molecular hydrogen (dotted).}
\label{Fig3}
\end{figure}

In the following sections we will assume that the GMCs 
in our sample are composed of 70\% molecular hydrogen
and 30\% helium. 
It is known (e.~g. Sunyaev \& Churazov 1996) 
that in the 
energy range considered scattering by the bound
electrons of 
molecular hydrogen and helium is important,
because for small scattering angles the differential
scattering cross-section per electron for Rayleigh scattering on 
molecular hydrogen and helium can become two times larger than the 
one for hydrogen atoms, which at these energies is identical 
to the Thomson differential cross-section. 
The angular dependence for Thomson scattering
is proportional to $(1+\cos^2\theta_\mathrm{sc})$,
where $\theta_\mathrm{sc}$ is the scattering angle of the photon,
which is approximately given by the angle $\alpha$ defined in Fig.~1.

In Fig.~3 we plot the scattered flux for a single cloud of
$10^6\,M_\odot$ composed of pure molecular hydrogen,
a mixture of 70\% molecular hydrogen and 30\% helium and pure
atomic hydrogen
as a function of the angle $\alpha$ ($\alpha=\pi$ corresponds to 
the case of forward scattering). We chose
a galactocentric distance and a distance to the observer of 5 kpc.
We used the differential scattering cross-section for scattering
by atomic, molecular hydrogen and helium as a function of photon energy 
and scattering angle as presented in Sunyaev et al.~(1999).

The scattered flux for the compositions including molecular hydrogen
are larger than the one for atomic hydrogen
because for small scattering angles ($\alpha\lsim\pi$) the two 
electrons present in molecular hydrogen and helium
scatter coherently. The scattered flux for the chosen 
mixture of molecular hydrogen and helium
is smaller than the one for pure molecular hydrogen because
in this case there are only .85 electrons per nucleon compared
to 1 electron per nucleon for pure hydrogen, atomic or molecular.
 
\section{Limits on the X-ray luminosity of Sgr A$^{*}$
         from Giant Molecular Clouds}
There exist several characteristic variability timescales 
for a supermassive black hole such as Sgr A$^{*}$. 
The shortest possible timescale 
is related to the linear size of the region, where
the bulk of energy is released 
in the accretion disk around a black hole
(Shakura \& Sunyaev 1973): 
$t_\mathrm{bh}\sim 10 R_\mathrm{g}/c \sim 300\,{\rm s}$ for Sgr A$^{*}$,
where $R_\mathrm{g}$ is the Schwarzschild radius of the black hole.
Tidal disruption events are believed to 
produce an Eddington level flare 
over a period $t_\mathrm{tidal} \sim {\rm years}$ (Rees 1988).
The lifetimes of active galactic nuclei (AGN) are believed
to be of the order of $10^{7}-10^{8}$ years.
The detailed temporal evolution of the luminosity of
Sgr A$^{*}$ is probably rather complex, possibly varying 
on all of the above timescales.

In this study it is our aim to constrain the past luminosity 
of Sgr A$^{*}$
from information about the reprocessing system, the spatially
distributed GMCs in the Galactic disk, and the output signal
at one point in time, the scattered flux observed today.
This is reminiscent of the methods used in reverberation mapping 
in AGN (see e.~g. Peterson 1993).
There one tries to gain information about
the geometry and kinematics of the emission-line region
(the reprocessing system)
by monitoring the continuum variations (the input) and the resulting
emission-line response (the output). 
Mathematically this is described by a transfer function,
which applied to our problem can be written as:
\begin{equation}
F_\mathrm{sc}(t)=\int_{-\infty}^{\infty}
\Psi(t')L_\mathrm{GC}(t-t')\,{\rm d}t'
\end{equation}
The scattered flux observed at a time t is the response 
of our system to activity of the GC in the past. 
In reverberation mapping in AGN
one determines the transfer function
using Fourier transforms of the input and the output over
a certain period of time.
In our case this is not possible because the scattered
flux or its upper limit is known only at one moment in 
time. 

If the cloud had no spatial extension, a ``point mass'',
the transfer function would be  
just a $\delta$-function
with an amplitude proportional to the mass of the cloud.
The transfer function for several clouds with different time
delays would be a sum of $\delta$-functions.
In these cases it would be possible to ascribe an unique 
X-ray luminosity of Sgr A$^{*}$ to an observed scattered flux.

Due to the spatial extension of a GMC
with an internal density distribution this 
$\delta$-function is smeared out in time.
This means that the scattered flux observed at one moment in time
depends upon the luminosity of the GC over a certain period of time
in the past. The correspondence between the observed
scattered flux and the luminosity of the GC is therefore
not one to one anymore. 

What one can do in this case to constrain the luminosity
of Sgr A$^{*}$  is to assume some input, a ``test function'', 
and compare the resulting output with the observations.
Should the output be inconsistent with the observed
flux then the initial input must be wrong.

We assume the temporal and spectral behaviour of
the luminosity of Sgr A$^{*}$ to be of the following form
\begin{eqnarray}
L_\mathrm{GC}(E,t)=
\cases{L_\mathrm{o}(E_\mathrm{o}/E)
& $t_\mathrm{on}$ $\leq$ $t$ $\leq$ $t_\mathrm{off}$\cr
0 & $\mathrm{else,}$\cr}
\end{eqnarray}
with $t_\mathrm{on}=t_\mathrm{o}-\Delta/2$ and
$t_\mathrm{off}=t_\mathrm{o}+\Delta/2$.
The choice of the spectral behaviour implies a photon index of 2,
which is representative of observed AGN spectra in the 4-10 keV range.
Note that here 
we define for the sake of symmetry $t_\mathrm{o}$ as the time half way through
the flare, whereas in Sec.~2.~1. $t_\mathrm{o}$ was defined as the
beginning of the flare.

Different inputs can produce the same output
because there is as mentioned above no one to one relation
between the luminosity and the scattered flux. 
It is therefore impossible to constrain the 
past luminosity at one moment in time.
The important quantity is thus actually the emitted
energy over a period $T$
\begin{equation}
Q=\int_{T} L_\mathrm{GC}(t')\,{\rm d}t',
\end{equation}
where $T$ is given by the minimum of the flare duration 
and the time for the ellipsoid to cross the cloud.
The limits we are deriving in the following sections
are therefore constraints upon a time averaged luminosity
of Sgr A$^{*}$, viz.~$\hat{L}_\mathrm{GC}\lsim L_\mathrm{max}$
with $\hat{L}_\mathrm{GC}=Q/T$.

\subsection{Response of a single cloud}
Observations of molecular tracers of H$_2$ 
(see e.~g. Blitz \& Williams 1999)
show that GMCs are far from being spherically 
symmetric objects. They are highly structured being made
out of clumps with high density variations from one point to another.
Due to their amorphous nature without a clear
central peak the density profile is often difficult to define
observationally. Another difficulty is that different tracers 
of H$_{2}$ turn optically thick at different densities. 
Nevertheless Williams et.~al (1995) were able to determine
a density distribution of the form $\rho\propto r^{-2}$
for dense clumps in the Rosette molecular cloud.
We therefore adopt this radial dependence 
as our profile A:
\begin{eqnarray}
\rho_\mathrm{cl}(r)=
\cases{\rho_{\mathrm{o}}
(r_\mathrm{o}/r)^2
& $r \leq r_\mathrm{cl}$\cr
0 & $r > r_\mathrm{cl}.$\cr}
\end{eqnarray}
The normalisation $\rho_\mathrm{o}$ is chosen in such a way, 
that the mass contained inside the radius of the cloud corresponds to 
the one given in SRBY87. The cutoff is required to obtain a finite mass.

Besides the response of the cloud with density profile A
we have also computed the response for a
different profile to study the influence of the
mass distribution inside a cloud upon our results. 
We choose as our profile B the following one:
\begin{equation}
\rho_\mathrm{cl}(r)=\frac{\rho_\mathrm{o}}
{(1+(r/r_\mathrm{o})^2)^2}.
\end{equation}
This profile has a constant core density up to a radius
of the order $r_\mathrm{o}$, which we choose to be
$0.1 r_\mathrm{cl}$. Most of the mass of this profile is 
provided by matter located at these radii.

Having specified the input (Eq.~18) and an internal 
density distribution for a cloud (Eqs.~20, 21)
we can integrate Eqs.~11 and 12 numerically to compute
the scattered flux produced by a cloud out of our sample. 
We chose cloud 116 from the SRBY87 sample (see Table 1)
and computed its response to Eddington flares of Sgr A$^{*}$ 
of durations of 1, 10, 100 and 300 years as a function of time. 
Fig.~4 displays the results of these computations.\footnote{Note that Fig.~4 
(and the following figures depicting a dependence 
upon $t-t_\mathrm{o}$) can be read in two ways:
\begin{enumerate}
\item Regarding $t$, the time of observation, as the independent 
      variable and keeping $t_\mathrm{o}$, the time half way through
      the flare, fixed (e.~g. $t_\mathrm{o}=-\delta$) the 
      different curves mark the scattered flux as a function
      of observing time t in response to an flare of Sgr A$^{*}$
      at $t_\mathrm{o}=-\delta$. One tracks the response of a 
      single flare as time evolves. 
\item Treating $t_\mathrm{o}$ as the independent variable and
      keeping $t$, the time of observation, fixed (e.~g. $t=0$) 
      the different curves mark the scattered flux observable
      at a time $t=0$ as a function of the time half way through
      the flare $t_\mathrm{o}$. At one moment in time one 
      observes the response of flares occuring at 
      different times $t_\mathrm{o}$. This case applies to
      what is done in this paper to constrain the past
      luminosity of Sgr A$^{*}$.
\end{enumerate}}

It shows that 
an Eddington flare of the GC lasting for one year  
$20\,764$ years ago, the time delay $\delta$ of cloud 116, would produce 
``today'' a scattered flux of about 
$2\times10^{-10}\,{\rm erg}\,{\rm cm}^{-2}\,{\rm s}^{-1}$ for profile
A and $4\times10^{-11}\,{\rm erg}\,{\rm cm}^{-2}\,{\rm s}^{-1}$ 
for profile B in the {\it ASCA}
field of view cloud 116 is located in. Since the
observational limit for this field of view is below these values, 
thus falsifying the initial assumption,
there could not have been an Eddington flare of Sgr A$^{*}$ of a
duration of one year or longer $20\,764$ years ago.
A one year flare occuring 100 years earlier or later can not be excluded
by cloud 116, because the scattered fluxes would be below the 
observational limits.   

\begin{figure}
\centering
\includegraphics[width=\columnwidth]{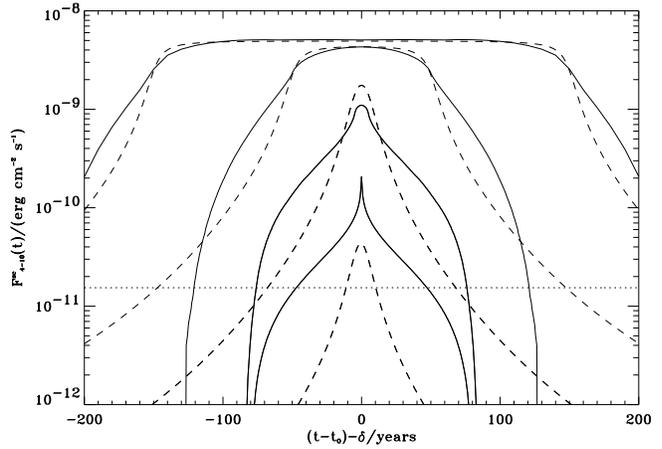}
 \caption{The scattered flux observable at a time t produced
          by a single cloud (number 116 
          in Table 1) in response
          to an Eddington flare of Sgr A$^{*}$ 
          of 1, 10, 100 and 300 years duration (from bottom to top)
          occuring around a time $t_\mathrm{o}$ 
          for two different density profiles of the cloud 
          given by Eq.~20 (solid lines)
          and Eq.~21 (dashed lines).
          The observational limit for the scattered flux produced
          in the field of view
          the cloud is located in is marked by a dotted line.
          The time delay $\delta$ for this cloud is  
          $20\,764$ years, the radius of the cloud is 13.4 pc
          and the time for the ellipsoid to cross the cloud 
          about 126 years, which corresponds to a cloud crossing velocity
          of $v_\mathrm{ell}\approx 0.8 c$ (for the other cloud parameters
          see Table 1).}
\label{Fig4}
\end{figure}

Because especially for short flares 
the shape of the lightcurve mainly depends upon how the mass
is distributed inside a cloud the responses for the density profiles
A and B differ.

The response of a cloud with a density profile A to 
the one year flare illustrates that a cloud whose mass is
centrally concentrated will produce a light curve with a 
very sharp maximum. 
This comes about because for short flare durations 
the space between the two ellipsoids covers only a
small part of the cloud. This illuminated region
sweeps along the cloud and produces a maximum flux 
when it covers the most dense, central part of the cloud.
Increasing the flare duration one encloses more
and more mass between the two ellipsoids, corresponding
to the beginning and the end of the flare. This increases
the scattered flux, which is proportional to the mass, 
and averages out density contrasts at different radii. 
The central spike produced by the one year 
flare therefore vanishes for longer flare durations.
The steep decline of the scattered flux observable for density 
profile A comes about because of the cutoff of the profile 
at $r=r_\mathrm{cl}$.

For density profile B, which has a constant density inside $r_\mathrm{o}$,
the main contribution to the total mass of the cloud 
is provided by matter at radii $r\sim r_\mathrm{o}$. 
Because the one year flare covers a region smaller than this size,
there is more mass probed by the one year flare in the central
region of profile A than profile B. The largest scattered flux
produced for a one year flare by a cloud with density profile A
is therefore larger than the one for a cloud with density profile B. 
For flares of longer duration larger parts of the cloud will be
covered and the fluxes will reach a similar level, because both
density profiles are normalised in such a way that the total mass
contained inside of a radius $r_\mathrm{cl}$ is equal for both
profiles. 

If the duration of the flare increases, the whole 
cloud will be covered at one point.
From this point on an increase of flare duration will no more
produce a larger flux, because all of the cloud is already
enclosed between the two ellipsoids.
This ``saturation'' time $t_\mathrm{cl}$, 
the flare duration for which the maximum
flux is achieved or equivalently the time for the ellipsoid to sweep
across the cloud, depends obviously on
the radius of the cloud but also on the position
of the cloud relative to the GC and the observer,
which determines time travel effects and the different 
velocities of the ellipsoid sweeping across the cloud. 
For example for the extreme case, when the cloud lies 
exactly behind
the source the time it takes an ellipsoid to cross the cloud
is $4\,r_\mathrm{cl}/c$.
On the contrary for the case when the cloud lies half way between
the GC and the observer at a small Galactic longitude, 
the time to sweep across the cloud
is of the order of
$(r_\mathrm{cl}/R_\mathrm{o})r_\mathrm{cl}/c$,
which is obviously much less than the value above.
Thus if the cloud is located closer to the observer
than the source of the flare, the whole cloud will
be scanned much quicker than the light crossing time
of the cloud. For the cloud behind the source
the temporal evolution will be slowest.
For the GMCs in our sample the values lie somewhere
between these two extreme cases.

We have computed lightcurves similar to Fig.~4 for all the GMCs in our
sample. Therewith we are able to estimate the saturation
time and the maximum scattered flux for them. 
The saturation time $t_\mathrm{cl}$
of a cloud and its radius yield an estimate for the velocity
of the ellipsoid sweeping across the GMC: 
$v_\mathrm{ell}\sim (2 r_\mathrm{cl}/c)/t_\mathrm{cl}$.
In Table 1, column 9 we have listed the saturation times
for twenty GMCs from our sample.
The cloud crossing velocities for these clouds lie in the range
$0.6 c\lsim v_\mathrm{ell} \lsim 3.2 c$.

The maximum scattered flux produced by a cloud yields an upper limit for
the luminosity of the GC via the following argument.
In our models we are computing the scattered flux in response
to an Eddington flare of Sgr A$^{*}$ some time ago, which enables us 
to estimate
the ratio $F_\mathrm{sc}/F_\mathrm{obs}$. 
Consistency with observations demand the scattered flux to be
smaller than the actually observed one: 
$F_\mathrm{sc}/F_\mathrm{obs}\leq1$.
Since the computed scattered flux just scales linearly with the assumed
input luminosity we can find
the luminosity of the GC for which this inequality is fulfilled:
\begin{equation}
L_\mathrm{GC} \lsim 
\frac{F_\mathrm{obs}}{F_\mathrm{sc}}L_\mathrm{Edd}.
\end{equation}
For twenty clouds selected from our sample we have listed
the scattered flux, the observed flux and the derived limit upon the
past luminosity of Sgr A$^{*}$ in columns 10-12 of Table 1.

If the duration of a flare increases
beyond $t_\mathrm{cl}$ 
then the flux will stay at a constant level
and the curves will extend to times 
$t-t_\mathrm{o}\approx\delta\pm\Delta/2$ as is illustrated
in Fig.~4 by the responses to the flare lasting 300 years. 
An even further increase of the flare duration will have
the result that
at one point several clouds along one line of sight
can be illuminated by one flare and their scattered fluxes begin to
overlap, producing a larger total flux.
This case will be discussed further below.

As mentioned in the beginning of this section observations
show that the above assumed spherically symmetric density profiles
are too simplified. The curves in Fig.~4 for a ``real'' GMC
will probably look quite differently. They will be noisier,
maybe show several peaks because of different density maxima
and will not display this symmetry. But, as we have shown, 
the choice
of the density distribution mainly determines the shape
of the lightcurves, the scattered flux as a function of time,
but not the level of the maximum scattered flux, which depends 
mainly on the total mass and the position of the cloud 
once it is fully covered by a flare.  
We are therefore confident as long as the CO observations
provide us with a good estimate of the total mass of a GMC
that our computations, especially of the maximum scattered flux
and the derived limit, are correct, albeit the mentioned
uncertainties about the density distribution inside a GMC.

\subsection{Response of one field of view}
\begin{figure}[t]
\centering
\includegraphics[width=\columnwidth]{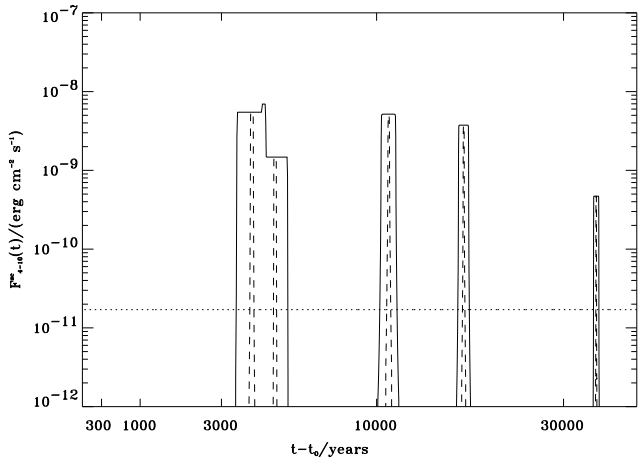}
 \caption{The scattered flux observable at a time t produced
          by five GMCs located in one {\it ASCA} field of view
          pointing at a Galactic
          longitude of $l=27.51^{\mathrm{o}}$ in response to an 
          Eddington flare of Sgr A$^{*}$ of 100 (dashed) and 
          1000 (solid) years duration
          occuring around a time $t_\mathrm{o}$.
          The observational limit for the scattered flux produced
          in this field of view is marked by the dotted line. Note how 
          the scattered fluxes of two GMCs begin to overlap for the
          1000 year flare because both are illuminated.}
\label{los}
\end{figure}
The {\it ASCA} field of view has a diameter of about
$40'$ and thus a size of $0.35\,{\rm deg}^2$.
This diameter corresponds to a linear scale of about
$100$ pc at the distance of the GC.
Fig.~2 shows that several clouds are located at 
similar Galactic longitudes. 
Their angular separations are often smaller than the
diameter of the {\it ASCA} field of view.
In most fields of view one finds therefore more than one cloud. 
Since we have a total of 96
GMCs distributed over 49 {\it ASCA} fields of view
there are about two clouds per field of view on average. 
For one of these fields of view ($l=27.51^{\mathrm{o}}$), 
in which five clouds are located, the maximum number of clouds
in one field of view, 
we plot the scattered flux as a function 
of time in Fig.~5.\footnote{The plots presented in 
this and the following subsection
have to cover a period of time 
of $40\,000$ years and also resolve flares on the timescale of years.
In order to be more illustrative we are plotting the total scattered 
flux as a function of a different time coordinate defined by:
\begin{eqnarray}
t\rightarrow\log\left(1+\frac{t}{t_\mathrm{s}}\right)
\nonumber
\end{eqnarray}
We have chosen $t_\mathrm{s}=3000$ years.
This choice ensures a linear behaviour for 
times small and a logarithmic one for times large compared 
to $t_\mathrm{s}$.}

Due to the different 
combinations of galactocentric and heliocentric distances
of these five clouds they have different time delays, i.~e.~ they
respond to activity of the GC at different times. For every cloud
we get a response similar to Fig.~4 at the time delay corresponding
to each cloud. 

Fig.~5 shows that all five clouds already have reached
their maximum flux for the 100 year flare. For this duration all
the clouds are already covered and their saturation time is therefore
smaller than 100 years.
For the 1000 year flare the scattered fluxes from two clouds start
to overlap. This happens when the duration of a flare
is longer than the time needed for the ellipsoid to
cross the distance between these two clouds.
Despite the fact that Fig.~5 restricts the luminosity of Sgr A$^{*}$
much more than Fig.~4 it is obvious that there are large time gaps
during the last $40\,000$ years where these 5 clouds are 
insensitive to the luminosity of Sgr A$^{*}$. 

Spatial overlapping of the clouds in the field of view
and absorption should not be a problem because the dense cores
with large column depths lie in different directions.
Certainly there should be further clouds of smaller size, 
below the detection threshold of the CO survey, present 
but these do provide only little scattered flux because of 
their much smaller masses. Since we have GMCs not only in one
field of view one can combine the limits from all fields of view,
which is done below.

\subsection{Response of all fields of view containing at least
               one GMC}

In total there are 49 {\it ASCA} fields of view covering at least one
cloud out of the SRBY87 sample.
The scattered flux produced in these fields of view will
be similar to the one shown in Fig.~5 with a peak for every cloud
roughly proportional to the mass of the cloud. The peaks are centered
around the time delay for the cloud and their temporal extent is 
of the order of $\Delta/2$, if the duration of the flare is longer than
the cloud crossing time. 

To combine the information from all 49 
fields of view we proceeded in the following way. 
We computed the expected scattered flux for each field of view
for a temporal grid covering the last
$40\,000$ years with a sufficient time resolution to resolve
the response of individual clouds. 
From these 49 lightcurves we select at each temporal grid point
the largest scattered flux produced in response to the assumed input.
As the observational limit at that time we choose
the limit for that {\it ASCA} field of view, in which 
the largest scattered flux is produced.
In Figs.~6-9 the result of this procedure is shown,
the temporal response of all fields of view
to an Eddington flare of Sgr A$^{*}$ of 
1, 10, 100 and 1000 years duration.

As one progresses from short to long flare durations 
more and more time is covered by the GMCs.
Increasing the flare duration by an order of magnitude 
results in an increase of the scattered flux by approximately the 
same factor (this factor depends upon the density profile) as long as the flare 
duration is smaller than the ``saturation time'' for a given
cloud. 

The first spike is produced by a GMC having a time delay
of about 335 years (cloud 014 in Table 1). Remarkable about
this cloud is that it attains its maximum scattered flux 
for a flare duration of only about 12 years. 
With its radius of $5.8$ pc one can estimate that 
the ellipsoid sweeps with a ``velocity'' of about $3.2\,c$
across the cloud. This very large velocity is the result
of the position of this GMC. It lies nearly half way
between the GC and the observer at an longitude of only
$10^{\mathrm{o}}$. In Sec.~3.~1. we showed that clouds
near the line between the GC and the observer can have 
very small cloud crossing times, because the motion
of the ellipsoid is highly superluminal.

The largest scattered flux is produced by a GMC with a time delay
of 669 years (cloud 003 in Table 1).
To make the maximum scattered flux produced by this cloud 
$(\sim 7\times10^{-8}\,{\rm erg}\,{\rm cm}^{-2}\,{\rm s}^{-1})$
consistent with the observational limits 
$(\sim 2\times10^{-11}\,{\rm erg}\,{\rm cm}^{-2}\,{\rm s}^{-1})$
the burst must have been $\sim3\times10^{-4}$ times below
the Eddington luminosity,
corresponding to a luminosity of Sgr A$^{*}$ of 
at most $\sim8\times10^{40}{\rm erg}\,{\rm s}^{-1}$.
For the smaller peaks one derives less tight limits (see also Table 1).

\begin{figure}[t]
 \centerline{\psfig{figure=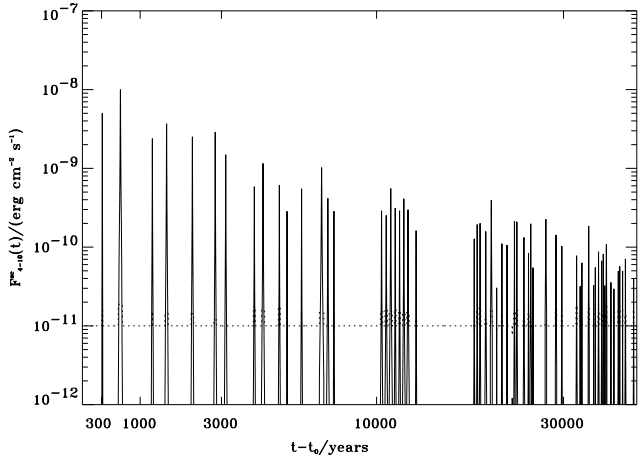, width=\columnwidth}}
 \caption{The maximum of the scattered flux observable at a time $t$
          in all the fields of view produced by all 96 GMCs from
          our sample in response to an Eddington 
          flare of Sgr A$^{*}$ of 1 year duration occuring 
          around a time $t_\mathrm{o}$. The observational limits 
          for the scattered flux are marked by a dotted line.} 
\end{figure}
\begin{figure}[t]
 \centerline{\psfig{figure=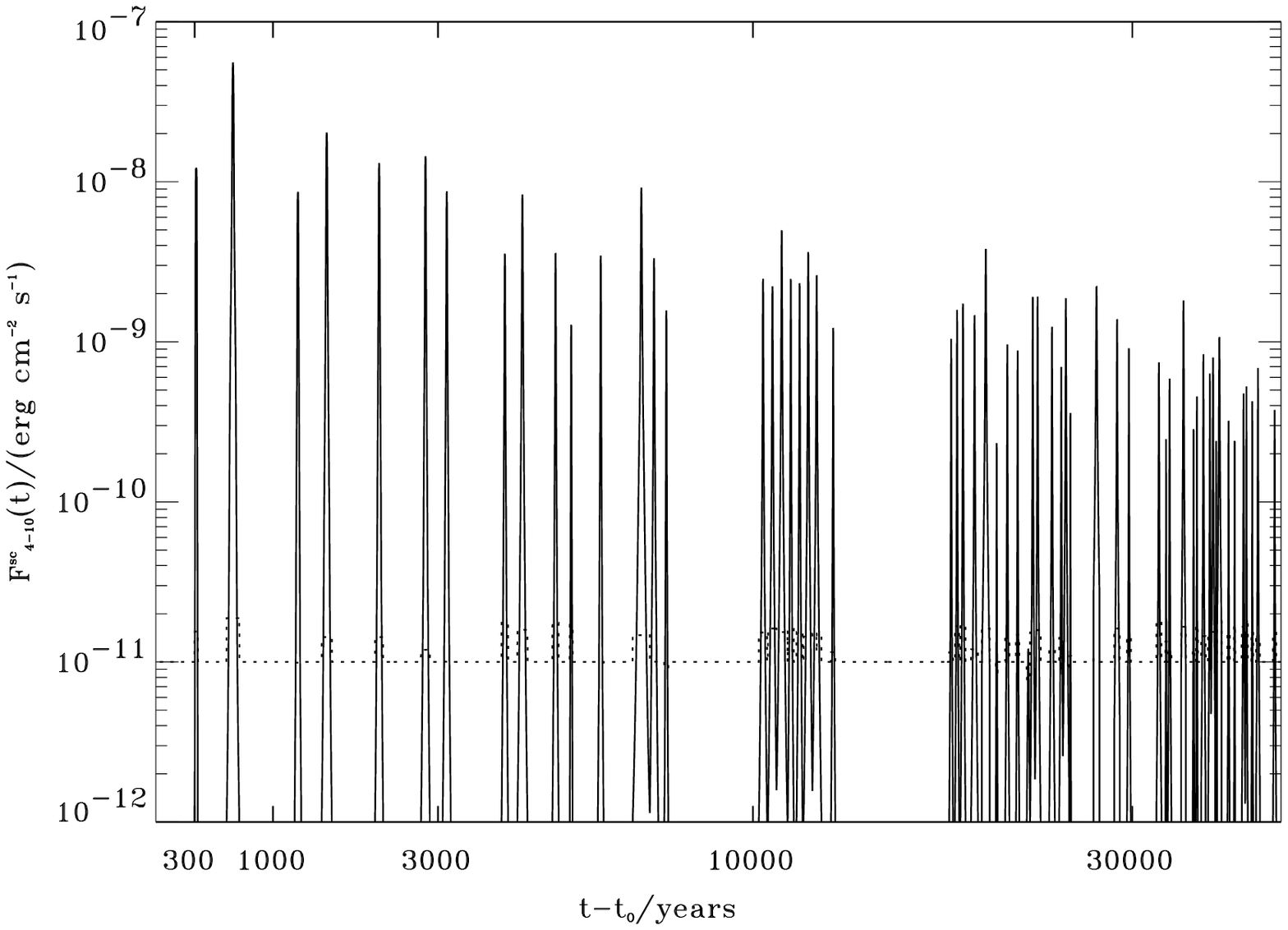, width=\columnwidth}}
 \caption{The maximum of the scattered flux observable at a time $t$
          in all the fields of view produced by all 96 GMCs from
          our sample in response to an Eddington 
          flare of Sgr A$^{*}$ of 10 years duration occuring 
          around a time $t_\mathrm{o}$. The observational limits 
          for the scattered flux are marked by a dotted line.} 
\end{figure}
\begin{figure}[t]
 \centerline{\psfig{figure=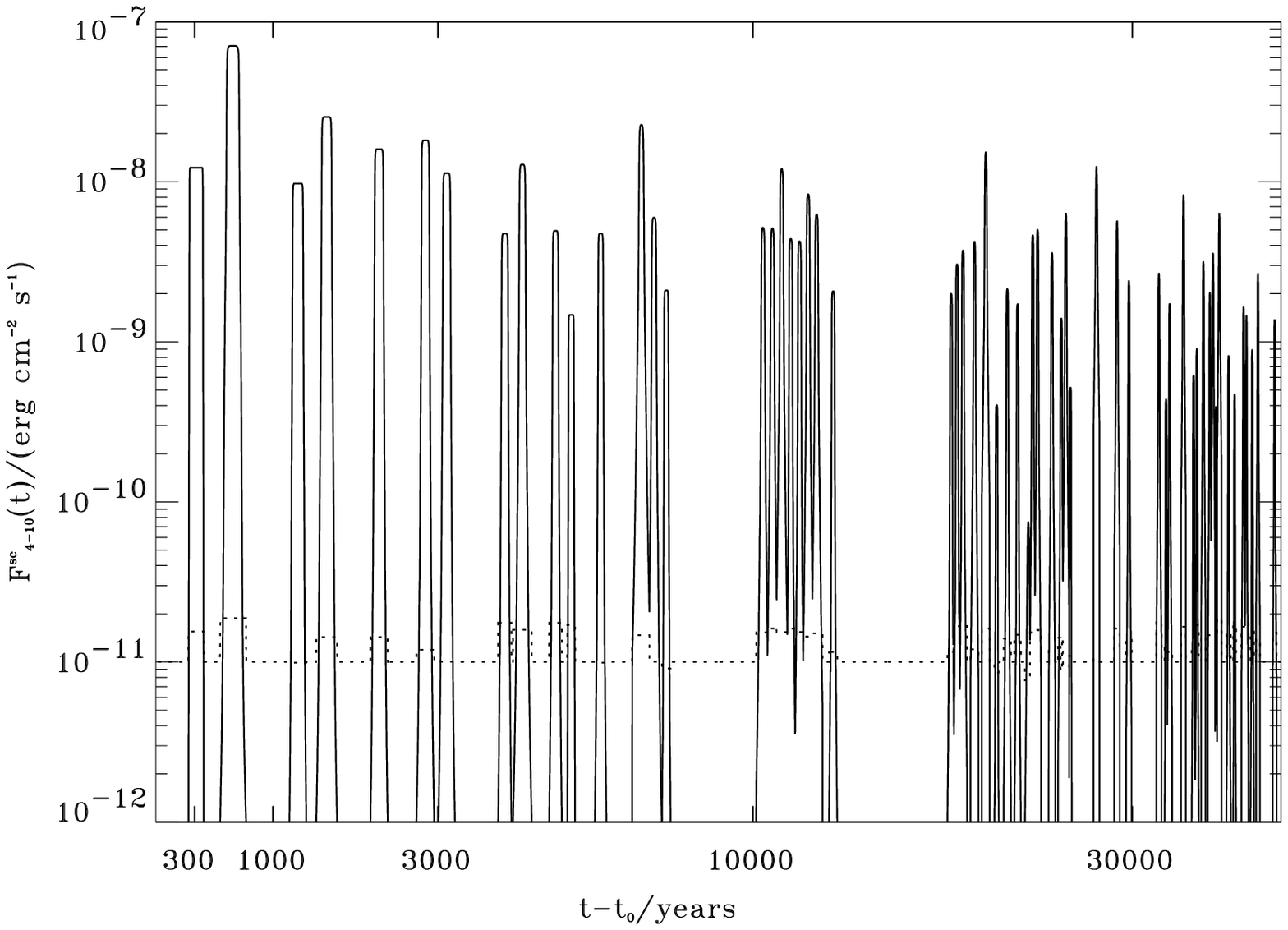, width=\columnwidth}}
 \caption{The maximum of the scattered flux observable at a time $t$
          in all the fields of view produced by all 96 GMCs from
          our sample in response to an Eddington 
          flare of Sgr A$^{*}$ of 100 years duration  occuring 
          around a time $t_\mathrm{o}$. The observational limits 
          for the scattered flux are marked by a dotted line.} 
\end{figure}
\begin{figure}[t]
 \centerline{\psfig{figure=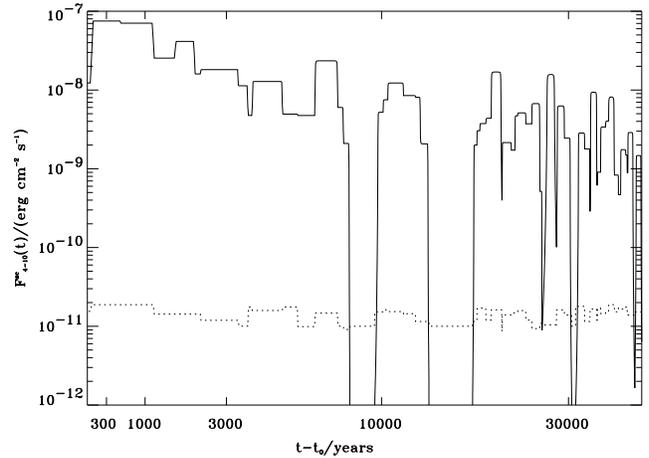, width=\columnwidth}}
 \caption{The maximum of the scattered flux observable at a time $t$
          in all the fields of view produced by all 96 GMCs from
          our sample in response to an Eddington 
          flare of Sgr A$^{*}$ of 1000 years duration  occuring 
          around a time $t_\mathrm{o}$. The observational limits 
          for the scattered flux are marked by a dotted line.}
\end{figure}

For short flare durations the flux from 
clouds with smaller time delays
is larger than the one from clouds with large time delays. 
Increasing the flare duration these clouds attain comparable
fluxes. This is mainly a consequence of the fact that 
the position of these clouds relative to the GC and
the observer, determines not only their time delay but also the 
time it takes the ellipsoid to cross the cloud. As we have seen
the velocity of the ellipsoid is smallest for clouds ``behind''
the GC, which are also the clouds with large time delays.  

Although the times when there is no observable scattered flux
shrink with increasing flare duration there are still some time 
gaps as shown in Fig.~9 even for an assumed flare duration
of 1000 years.
The two most pronounced time gaps lie around about 8000 and 
$14\,000$ years.
It is obvious from the trend in Figs.~6-9 that by further 
increasing the assumed flare duration at one point 
the whole period of the last $40\,000$ years will be covered.
This happens for flare durations longer than about 3000 years.
This means that we can
exclude a flare lasting longer than 3000 years of a luminosity
larger than a few $10^{42}\,{\rm erg}\,{\rm s}^{-1}$ to have happened
during the last $40\,000$ years. 
A similar conclusion can be drawn from Fig.~10.
Here we tried to address the question of what is the 
chance that there was a strong flare of duration
$\Delta$ during the last $40\,000$ years unnoticed by us?

Fig.~10 shows for the clouds from the SRBY87 sample
the fraction of time the scattered flux
was larger than the observational limits  
during the last $40\,000$ years for 
different flare durations.
This quantity should be an approximate estimate of the 
probability to detect
a flare as a function of 
flare duration $\Delta$.
The curves should be continuous but we computed them only
for several durations of the flare.
It is obvious that this simple 
estimate does not take into account the level of the
scattered flux above the 
observational limits. 
From this picture we conclude that an Eddington level 
flare of duration longer than about 300 years 
occurring any time during
the last $40\,000$ years is unlikely
and one of duration longer than 3000 years 
can be excluded.  
Since as mentioned above we only were able to use 
the data from one Galactic quadrant the values of 
Fig.~10 actually should be regarded as lower limits 
to the probability.
\begin{figure}
\centering
\includegraphics[width=\columnwidth]{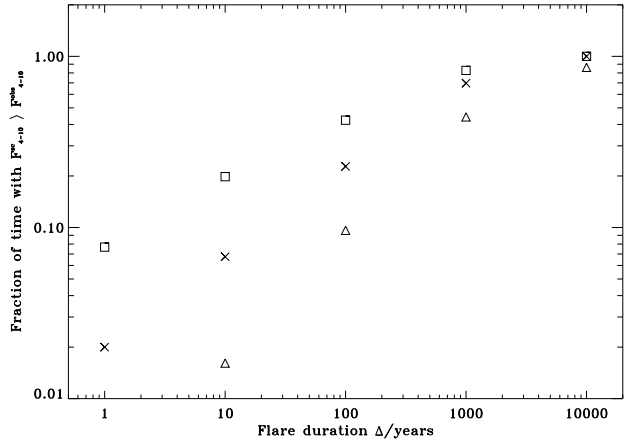}
 \caption{The fraction of time the scattered flux is 
          larger than the observed upper limits during
          the last $40\,000$ years as a function
          of the duration of the flare for different luminosities of
          Sgr A$^{*}$: 
          $10^{40}\,{\rm erg\, s}^{-1}$(triangles), 
          $10^{42}\,{\rm erg\, s}^{-1}$(crosses) and  
          $3\times10^{44}\,{\rm erg\, s}^{-1}$(squares).
          This quantity is an approximate estimate of the
          probability to detect a flare of duration $\Delta$
          occuring during the last $40\,000$ years.}
\label{prob}
\end{figure}

\section{Limits on the X-ray luminosity of Sgr A$^{*}$ 
         from the interstellar HI distribution}
Another major component of the ISM of the Milky Way 
is neutral, atomic hydrogen.
Its total mass is about three times larger than that of 
molecular hydrogen. This is due to the larger extent
of the HI disk. 

As shown in the previous section our sample
of GMCs is insensitive to the activity of the 
GC during the last $40\,000$ years
for a few time windows, in particular two periods
of 2000 and 4000 years duration of about 8000 and $14\,000$ years ago
for flares of about 1000 years duration.
For shorter flare durations these temporal gaps become larger 
and more numerous.  
They are as mentioned in Sec.~1 a result of the concentration of GMCs towards the
molecular ring, their finite number in the Galactic disk,
their finite size of about 10 pc and the probably 
the fact that we are missing the necessary
data about GMCs located in the fourth Galactic quadrant. 
In this respect the 
more continuous HI distribution is of greater value.
It provides less tight limits but extends further back into the past
without large gaps in the temporal coverage.

Interstellar neutral hydrogen is concentrated in clouds, 
which are much less massive $(\lsim 10^{4}\, M_\odot)$ than GMCs. 
Their total number is thus much larger.
In the model of the Galactic ISM
of McKee \& Ostriker (1977) the mean separation between
individual HI clouds $\lambda$ is of the order of 88 pc. 
Applied to our problem there exist two obvious limits
for the response of these clouds distributed in the Galactic disk
to a flare of Sgr A$^{*}$ depending on its duration:

\begin{list}{}{}
\item[a)] 
If the flare duration $\Delta$ is longer than the 
time for the ellipsoid to travel the mean distance 
between HI clouds, $\lambda/c\approx290$ years,  
the flare scatters
on a quasi continuous background medium. 
Because the gas is optically thin 
all the mass illuminated by the flare, the mass
enclosed between the two ellipsoids, contributes to the
scattered flux. 
\item[b)]
For a flare duration shorter than the time it takes 
the ellipsoid to travel across the mean
cloud separation we are dealing with
individual clouds and the problem becomes more complex,
in which each HI cloud gives some contribution
to the total scattered flux.
\end{list}

The HI distribution is rather thin with a scale height
of about 100 pc (e.g.~Dickey \& Lockman 1990)
for galactocentric distances smaller than 
$R_\mathrm{o}$. Beyond the solar circle
the HI scale height seems to increase due to 
a possible warp of the Galactic disk. 
For distances larger than 
$\sim R_\mathrm{o}$ most of the disk 
is practically inside the {\it ASCA} field of view 
of $40'$. For our purposes it is therefore valid to 
neglect the distribution of HI with height above the disk
and to assume the HI to be concentrated in the Galactic disk.
Since the HI mass surface 
density is approximately constant out to a radius of 16 kpc, 
we assume the following profile:
\begin{eqnarray}
\Sigma_\mathrm{HI}(R)=
\cases{4\,M_\odot\,{\rm pc}^{-2}
& $R \leq R_\mathrm{max}$\cr
0 & $R > R_\mathrm{max},$\cr}
\end{eqnarray}
with $R_\mathrm{max}=16\,{\rm kpc}$.
In this section we want to compute how the HI distributed in our Galaxy according  
to this profile responds to different types of activity of Sgr A$^{*}$.
The forms of activity we have in mind are:
\begin{itemize}
\item
{\bf Switch off}: Sgr A$^{*}$ was active at a level $L_\mathrm{Edd}$
up to a time $t_\mathrm{o}$
and then switched off.
\footnote{
Note that in this and the following section
$t_\mathrm{o}$ marks the end of the assumed activity.}
\item
{\bf Flare}: A flare lasting for a duration $\Delta$ at a level $L_\mathrm{Edd}$
ended at a time $t_\mathrm{o}$.
\end{itemize}
For the Milky Way the switch off scenario is equivalent
to a ``flare'' with a duration longer than about $110\,000$ years
because the whole disk out to 16 kpc is illuminated in this case. 

For the mass surface density profile given by Eq.~23
the mass of hydrogen distributed along the {\it ASCA} 
beam pointing at a longitude of
$10^{\mathrm{o}}$ is of the order of 
$2\times 10^{7}\,M_{\odot}$, which is about an order of magnitude
larger than the most massive GMCs in our sample.
It is therefore of no surprise that if all of this gas is illuminated
the scattered flux will be substantial. 
The maximal optical depth for Thomson scattering along an {\it ASCA}
beam is of the order of $5\times10^{-2}$. 
The main photoelectric absorption by heavy elements
should be caused by GMCs along the way. 
One should therefore look for an
{\it ASCA} field not ``contaminated'' 
by a massive GMC. 

\begin{figure}
\centering
\includegraphics[width=\columnwidth]{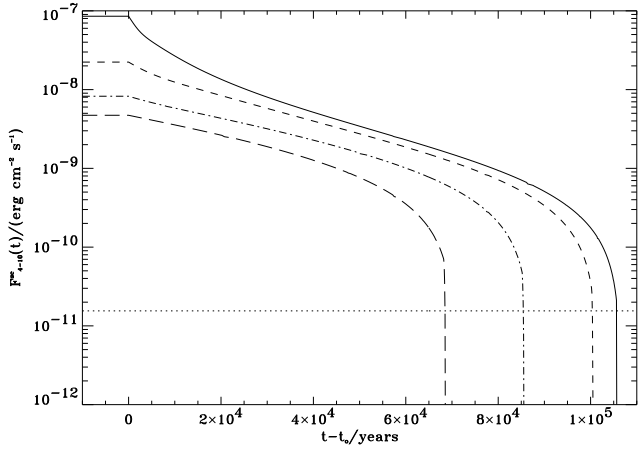}
 \caption{The scattered flux observable at a time t 
          produced by the Galactic HI
          distributed along one of the {\it ASCA} fields of view 
          in response to Sgr A$^{*}$
          switching off from Eddington luminosity
          at a time $t_\mathrm{o}$. 
          The different
          lines mark fields of view pointing at different
          Galactic longitudes: $\pm 10^{\mathrm{o}}$ (solid), 
          $\pm 30^{\mathrm{o}}$ (dashed), $\pm 60^{\mathrm{o}}$
          (dot-dashed) and $\pm 90^{\mathrm{o}}$ (long dashed). The 
          observational limit for the flux in the {\it ASCA} field of view
          pointing at $l=10^{\mathrm{o}}$ is marked
          by the dotted line.}
\label{back}
\end{figure}

In Fig.~11 we plot the scattered flux produced at a time $t$ 
by the HI distributed along {\it ASCA} beams pointing at four
different Galactic longitudes in response to a switch off 
of Sgr A$^{*}$ at a time $t_\mathrm{o}$.
The further away one looks from the GC 
the less mass will be located inside the field of view.
This is the main reason for the differences of the scattered 
fluxes for different longitudes at times before the switch off.
The cutoff of the curves at late times
is due to the ellipsoid
leaving the disk. For larger Galactic longitudes where the line of 
sight through the disk is smaller this occurs at earlier times, e.~g.
for $90^{\mathrm{o}}$ this happens after about $70\,000$ years
whereas for $10^{\mathrm{o}}$ after $100\,000$ years.

\begin{figure}
\centering
\includegraphics[width=\columnwidth]{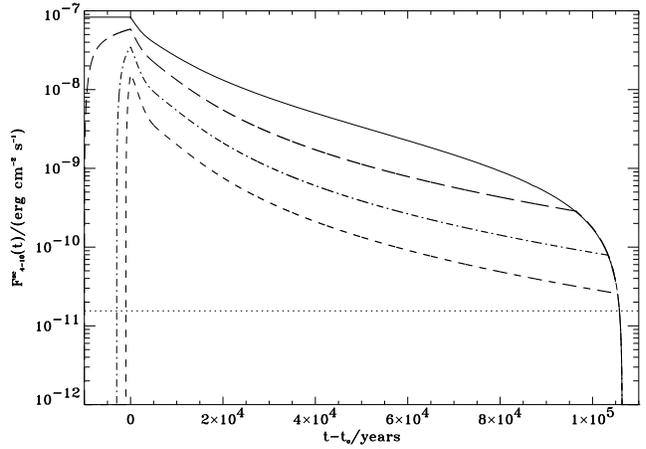}
 \caption{The scattered flux observable at a time t
          produced by the Galactic HI
          distributed along one of the {\it ASCA} fields of view 
          pointing at a Galactic longitude $l=10^{\mathrm{o}}$
          in response to Eddington flares of Sgr A$^{*}$
          of different durations ending at a time 
          $t_\mathrm{o}$:
          switch off (solid), $10\,000$ years (long dashed), 
          $3000$ years (dot-dashed) and 1000 years (dashed).
          The observational limit for the flux in the {\it ASCA}
          field of view pointing at $10^{\mathrm{o}}$ is marked
          by the dotted line.}
\label{back}
\end{figure}

Fig.~12 shows the same quantities plotted for one Galactic
longitude but for different durations of the flare. 
It is obvious
that a switch off of Sgr A$^{*}$
from Eddington level less than about $100\,000$ years ago
is inconsistent with the observational limits. 
For a luminosity typical of AGN hosted by a spiral galaxy 
of $10^{43}\,{\rm erg}\,{\rm s}^{-1}$ a switch off could not
have happened less than about $80\,000$ years ago. 
Remarkable about this picture is
that even a flare of a duration of 1000 years happening
$100\,000$ years ago would produce a scattered flux larger than 
currently observed upper limits. A similar conclusion can
be drawn from the following two figures.

Fig.~13 shows the dependence of the scattered flux in an {\it ASCA}
field of view at different times after the end of the activity in
the longitude range $-60^{\mathrm{o}}\leq l \leq 60^{\mathrm{o}}$. 
Due to our assumed axially symmetric mass surface 
density profile 
the curves are obviously symmetric in the Galactic plane 
about the direction towards the GC.
At small longitudes the scattered flux is large 
because one looks through a lot of gas
and also the distance of the gas to Sgr A$^{*}$ is smaller
than for larger longitudes.
The scattered fluxes are rather large, because for the switch off case 
the whole disk is scattering at early times after the demise of the 
nuclear activity. The cutoff observable for the scattered flux
$100\,000$ years after the switch off at Galactic longitudes 
$|l|\gsim 35^{\mathrm{o}}$ is again a result of the ellipsoid
growing beyond the outer edge of the HI disk and therefore 
leaving no illuminated matter along that direction. 

For a shorter flare the flux will be obviously less, because only the
illuminated parts of the disk will scatter the incident radiation. 
In Fig.~14 we plot the
expected flux for a flare of 300 years duration. As argued above
flares of this or a longer duration should scatter on a ``continuous'' 
HI background. 
\begin{figure}[t]
\centering
\includegraphics[width=\columnwidth]{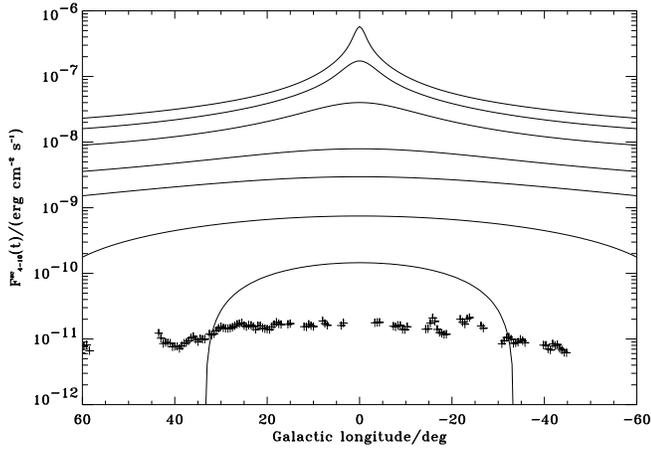}
 \caption{The scattered flux produced by the Galactic HI
          distributed along one {\it ASCA} field of view 
          in response to Sgr A$^{*}$
          switching off from Eddington luminosity
          at a time $t_\mathrm{o}$ as a function
          of Galactic longitude $l$. The different lines 
          mark the scattered flux observable at
          different times after the
          switch off of Sgr A$^{*}$: 1000, 3000, 
          $10\,000$, $30\,000$, $50\,000$, $80\,000$ 
          and $100\,000$ years
          (from top to bottom).
          The observational {\it ASCA}
          limits are marked by crosses.}
\label{back}
\end{figure}

The dependence of the flux scattered by the Galactic HI 
upon Galactic longitude
in response to a flare with a suitable choice
of duration, time elapsed since the fading of the flare  
and luminosity is similar to the 
X-ray emission of the Galactic ridge as observed by {\it ASCA}. 
This scenario might therefore be a possible explanation 
for the hard component of the X-ray emission of the Galactic ridge.
Under the assumption that the flare of Sgr A$^{*}$ had a hard enough
spectrum the scattered emission might contribute to 
the hard component of the emission of the Galactic 
ridge detected above 10 keV by {\it RXTE} (Valinia \& Marshall 1998) and
{\it OSSE} up to energies of 600 keV (Skibo et
al.~1997). There is one crucial test for this model. If we really
observe scattered radiation from Sgr A$^{*}$
then a narrow iron K$_{\alpha}$-line 
at $6.4$ keV must be present with an equivalent width of the order
of 1 keV (Vainshtein \& Sunyaev 1990).

The detection of emission in the 6.7 keV line from
the Galactic ridge by {\it Tenma} was reported by Koyama et al.~(1986).
This fact shows that the upper limits from {\it ASCA} we are using
are coming from the detection of the non scattered component. The
detection of the neutral iron K$_{\alpha}$-line 
intensity in the same direction 
might give us much stronger limits than presented in this paper.
{\it XMM-Newton} is certainly able to separate the 
6.4 and 6.7 keV lines much better than {\it Tenma}.
Therefore we will get this information in the coming years and
will be able to improve on the results of this paper. 

\begin{figure}[t]
\centering
\includegraphics[width=\columnwidth]{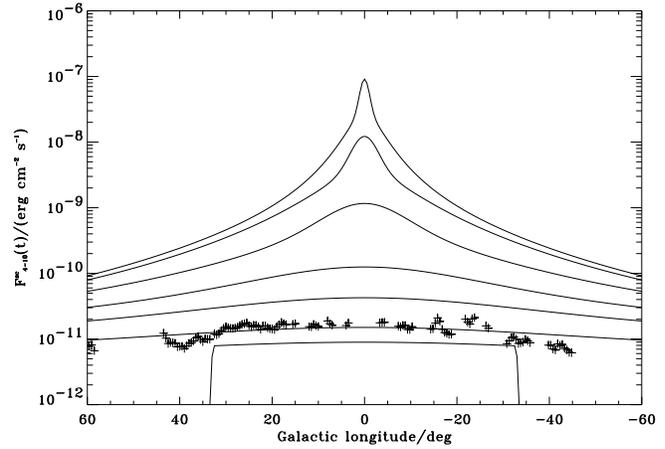}
 \caption{The scattered flux produced by the Galactic HI
          distributed along one {\it ASCA} field of view 
          in response to an Eddington flare of Sgr A$^{*}$ of 300 years
          duration
          ending at a time $t_\mathrm{o}$ as a function
          of Galactic longitude $l$. The different lines 
          mark the scattered flux observable at
          different times after the
          end of the flare: 1000, 3000, 
          $10\,000$, $30\,000$, $50\,000$, $80\,000$ 
          and $100\,000$ years (from top to bottom). 
          The observational {\it ASCA}
          limits are marked by crosses. It is possible
          for the flare to reach large longitudes because
          it moves with highly superluminal motion at early times.}
\label{back}
\end{figure}

Obviously spiral arms or other large scale structures,
a warp of the Galactic disk, will change the simple picture
described above. It will lead to increased scattering 
for some directions, e.~g. for large 
column densities along spiral arms,
and thereby provide additional information
about these deviations from a constant disk.
If for example the density contrast of HI inside and outside of
spiral arms is substantial one would get an additional
time dependence as the ellipsoid scans across spiral arms
and the regions between them.
Nevertheless in this paper we decided to assume the 
simplified distribution given by Eq.~23.

\section{The Milky Way at extragalactic distances}
It is well known that other spiral galaxies contain 
an ISM similar to the one of the Milky Way. 
CO and 21 cm observations have revealed the 
spatial distribution of H$_{2}$ and HI in many 
nearby spiral and irregular galaxies.
M31 for example, containing a supermassive black hole 
of about $4.5\times10^{7}\,M_{\odot}$
in its center (see e.~g. Kormendy \& Gebhardt 2001), 
has a HI mass similar to the one of  
the Milky Way, which is very concentrated in a ring
at about 12 kpc from the nucleus (Sofue \& Kato 1981).
Although its H$_{2}$ mass is less than the one of the 
Milky Way it is also located in a ring at about 
the same radius (Nieten 2001; Wielebinski priv.~comm.).
Since there have been and will be {\it Chandra} 
(Garcia et al.~2000)
and {\it XMM-Newton} (Shirey et al.~2001) 
observations of this object, M31 might be a good target to apply to 
the methods and results presented in this section. 
Many other nearby spiral galaxies have been observed by
{\it Chandra} already, e.~g. M100 (Kaaret 2001), M101 (Pence et
al.~2001), NGC 1068 (Young et al.~2001) and 
NGC 4151 (Yang et al.~2001) to name a few.

Many of these spiral galaxies
also seem to harbour a supermassive black hole at their centers 
(Kormendy \& Gebhardt 2001)
with some of them demonstrating AGN activity.
Spiral galaxies are believed to be the host galaxies
of certain classes of AGN, in particular 
Seyfert I and II galaxies. The luminosities of their nuclei are 
typically of the order of $10^{42}-10^{44}\,{\rm erg}\,{\rm s}^{-1}$.
It is therefore an interesting question to ask 
how interstellar gas responds to activity of the AGN
and what observational signs it should produce when
the AGN switches off.
We want to address these questions using the methods 
presented in the previous sections. 

To demonstrate solely the effects of scattering by a
gas disk we decided
not to model a specific galaxy with an observed 
density distribution of its neutral gas but rather
to show how the scattered emission of the Milky Way
would look like at extragalactic distances.
We therefore assume the same simple mass surface 
density profile as in the previous section (Eq.~23).

\subsection{Scattered X-ray surface brightness}
\begin{figure}
\centering
\includegraphics[width=\columnwidth]{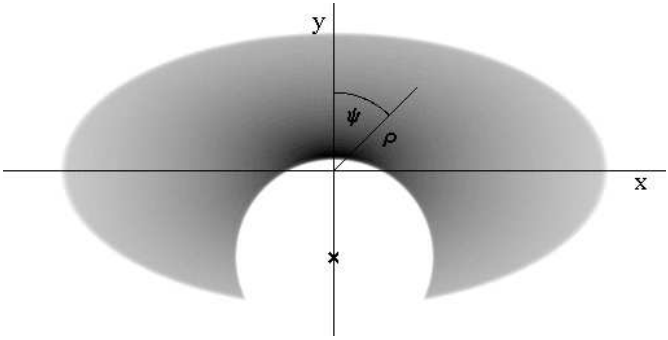}
  \caption{The surface brightness of the scattered emission in the plane
           of the sky (specified by $x$ and $y$ or $\rho$ and $\psi$)
           produced by a fully illuminated gas disk of constant 
           mass surface density $10\,000$ years
           after the central source has turned off.
           The inclination angle of the disk is $i=60^{\mathrm{o}}$.
           The physical radius of the disk is 16 kpc.
           The center of the circle (marked
           by a cross) which marks the boundary between the 
           illuminated and not illuminated
           parts of the disk moves with a velocity 
           of $1.73c$ in the direction given by $\psi=\pi$. 
           Its
           radius has a size of about 6 kpc and grows
           with time at a rate $2c$.}
\label{andromeda}
\end{figure}
If a galaxy is close enough to be resolved by the
{\it Chandra} X-ray 
telescope then it is possible to map its scattered 
surface brightness as a function of position on the sky. 
The basic formula to compute this quantity along 
a given line of sight is Eq.~11.
In this section we are using the paraboloid approximation
(see Sunyaev \& Churazov 1998 or Sec.~1) 
to compute the integration boundaries $s_\mathrm{min}(t)$ 
and $s_\mathrm{max}(t)$. The paraboloid approximation is valid in
this case given the large distances of extragalactic objects. 

In Fig.~15 we plot the surface brightness of the scattered emission 
produced by a gas disk with a central, 
isotropic radiation source of luminosity 
$L_\mathrm{agn}=10^{44}\,{\rm erg}\,{\rm s}^{-1}$ 
that turned off $10\,000$ years ago.
The disk has an inclination of $i=60^{\mathrm{o}}$,
with $i$ defined as the angle between the line of sight and the normal
direction of the disk.
The figure shows that a part of the disk is not bright, 
because it is not illuminated anymore\footnote{Note 
that the ``negative'' of Fig.~15 
corresponds to the case $10\,000$ 
years after a switch on of the AGN.}.
The boundary of this region, which is the 
projection of the intersection of the paraboloid at a time
$t-t_\mathrm{o}$ 
with the disk onto the plane of the sky, is a circle.
As time progresses this boundary moves 
radially outwards from the nucleus with an apparent velocity 
\begin{equation}
\dot \rho(\psi)=
\frac{c \cos i}{1+\cos\psi\sin i}.
\end{equation}
This velocity depends upon the position angle $\psi$, which is defined in Fig.~15. 
The maximum velocity of $c \cos i/(1-\sin i)$  
is given for $\psi=\pi$ 
and the minimum velocity of $c \cos i/(1+\sin i)$ for $\psi=0$.
This ``motion'' has the effect that 
the center of the circle (marked by a cross in Fig.~15)
moves in the direction $\psi=\pi$ while it is growing in size. 
The velocity of its center is given by   
$v_\mathrm{o}=(\dot \rho(\pi)-\dot \rho(0))/2=c\,\tan i$,
which is superluminal for inclination angles larger than 
$45^{\mathrm{o}}$.
For an inclination of $i=60^{\mathrm{o}}$ this velocity is 
$v_\mathrm{o}\approx1.73c$.
The radius of the circle grows with time as
$c\,t/\cos i$. 

The motion of this circle and the boundary between the illuminated
and not illuminated parts of the disk is an example of apparent 
superluminal motion produced by scattering on suitable shaped 
screens. This subject has been discussed in the literature 
by e.~g. Blandford et al.~(1977).

\begin{figure}
\centering
\includegraphics[width=\columnwidth]{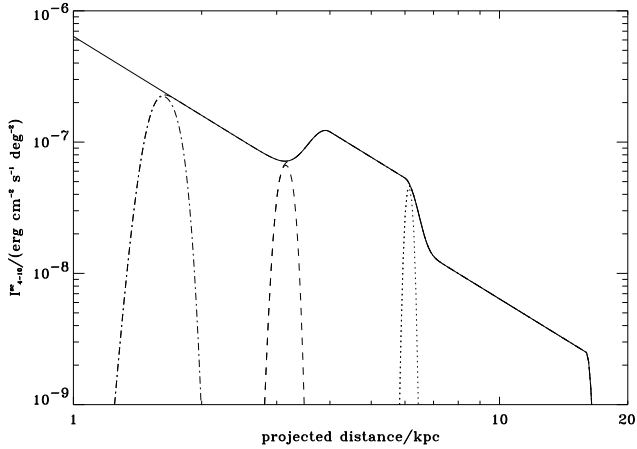}
  \caption{The scattered X-ray surface brightness as a function of projected
           distance from the nucleus along a fixed position angle of
           $\psi=\pi/2$. The solid line 
           marks the surface
           brightness for the fully illuminated disk. The 
           other lines mark the switch off (Fig.~15) 
           and the 1000 year flare case (Fig.~19) 
           5000 (dot-dashed), $10\,000$ (dashed) 
           and $20\,000$ (dotted) years after the end of
           the nuclear activity. 
           The switch off case corresponds to the curves which
           show no brightness below a certain distance and then
           join on to the solution for the fully illuminated disk.
           The increase in surface brightness 
           between 4 and 6 kpc is
           due to the molecular ring which is located at these radii.
           The $\rho^{-2}$ dependence described by Eq.~25 is visible.}
 \label{andromeda}
\end{figure}

For a face on disk ($i=0^{\mathrm{o}}$) with a constant mass surface 
density it is easy to see from Eq.~11 that the surface brightness 
drops with distance from the nucleus as $\propto \rho^{-2}$.
This is still valid in the case of an inclined disk.
In the limit of a uniform disk with a constant mass surface
density restricted by $r=r_\mathrm{max}$ 
one can derive an analytic solution for the surface brightness 
of the illuminated parts of a disk with inclination $i$.
It is necessary to express
the factor $(1+\cos^2\theta_\mathrm{sc})/r^2$ appearing
in Eq.~11, where $\theta_\mathrm{sc}$ is the scattering angle and
$r$ the physical distance from the nucleus, as a function
of the projected radius $\rho$, the position angle $\psi$ and
the inclination angle $i$ which yields
\begin{equation}
I_{\nu}=
\,
\frac{\Sigma_\mathrm{HI}}{\cos i}
\,
\frac{L_\mathrm{agn}}{4 \pi \rho^2}
\,
\frac{3\sigma_\mathrm{T}}{16 \pi}
\,
\frac{1+2\cos^2 \psi \tan^2 i}{(1+\cos^2 \psi \tan^2 i)^2}.
\end{equation}    

As visible from Fig.~15, beyond a certain position angle the scattered
surface brightness cuts off because these parts of the disk are not
illuminated anymore by the central source. This angle is given by: 
\begin{equation}
\psi_\mathrm{max}=\arccos 
\left(
\frac{c^2t^2-\rho^2}
     {2\,c t \rho \tan i}
\right).
\end{equation}

\begin{figure}
\centering
\includegraphics[width=\columnwidth]{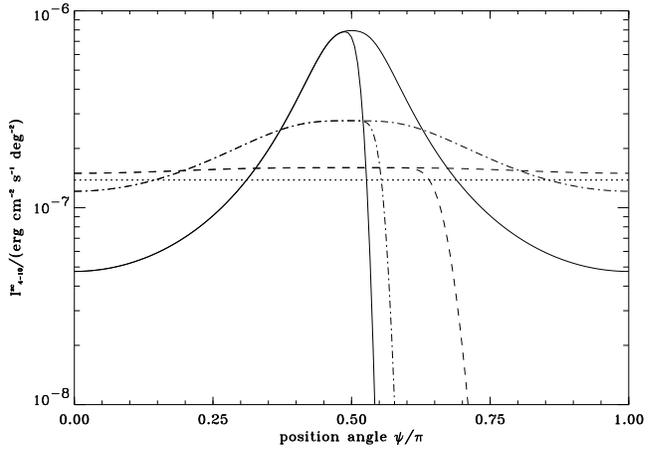}
 \caption{The scattered X-ray surface brightness as a function of position angle at
          a fixed projected distance of 2 kpc from the nucleus.
          The different lines mark different inclination angles
          of the disk for the fully illuminated disk and 5000 
          years after switch off: $i=0^{\mathrm{o}}$ (dotted),
          $i=30^{\mathrm{o}}$ (dashed), $i=60^{\mathrm{o}}$ (dash-dotted)
          and $i=80^{\mathrm{o}}$ (solid). The switch off leads to to
          disappearance of the illuminated region at large position angles.}
\label{andromeda}
\end{figure}

Besides this more qualitative picture we present 
in Figs.~16, 17 and 18 the surface brightness along several 
cuts of the partly illuminated disk shown in Fig.~15.
These figures confirm the validity of the above formulae, Eqs.~25 and 26.

In Fig.~16 we plot the scattered surface brightness as a function of
projected radius $\rho$ for a fixed position angle of $\psi=\pi/2$.
The curve for the fully illuminated disk shows the $\rho^{-2}$
dependence of the surface brightness given by Eq.~25. 
Note that in this computation 
we have included a molecular ring with a two times larger mass surface
density between about 4 and 6 kpc, which causes the bump
between these radii. The surface brightness cuts off beyond the maximal 
disk radius of 16 kpc. 
Furthermore plotted in this figure are the surface brightness
for the switch off (full picture
given in Fig.~15)
and the flare scenario (full picture given in Fig.~19) at different
times after the end of the nuclear activity. 
The curves showing the surface brightness for the the flare
scenario move with a velocity of about $c$ to larger projected 
distances as given by Eq.~24 for a position angle of $\psi=\pi/2$.

Figs.~17 and 18 show the scattered surface brightness 
at a fixed projected distance of $\rho=2\,{\rm kpc}$ 
from the nucleus as a function of position angle.
For clarity no molecular ring was assumed in these two 
plots.

In Fig.~17 we plot the surface brightness of the scattered emission
for different inclinations of the disk for the fully illuminated
and the switch off case.  
The brightness contrast for large inclinations
of the disk mainly comes about because the same projected distance at
position angles $\psi=0$ and $\psi=\pi/2$ corresponds to physical
distances from the nucleus differing by a factor of $(\cos i)^{-1}$. 
Furthermore the average scattering angle for material at a position
angle of $\psi=0$ is $\theta_\mathrm{sc}\approx\pi/2+i$ 
compared to $\theta_\mathrm{sc}\approx\pi/2$ 
for a position angle of $\psi=\pi/2$.  
The angular dependence of the curves presented in Fig.~17 
is well described by Eq.~25.

Fig.~18 shows the same quantities plotted for one disk 
with a large inclination of $i=80^{\mathrm{o}}$ at different
times after the switch off of the central source.
With increasing time the not illuminated part of the disk
``works'' its way towards smaller position angles. The angle where
the brightness cuts off is given by Eq.~26.

\begin{figure}
\centering
\includegraphics[width=\columnwidth]{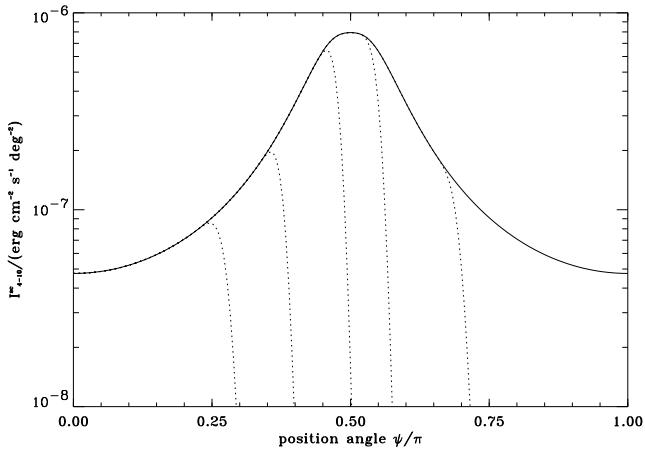}
 \caption{The scattered X-ray surface brightness as a function of position angle at
          a fixed projected distance of 2 kpc from the nucleus.
          The inclination of the disk is $i=80^{\mathrm{o}}$. 
          We plot the surface brightness for the fully illuminated 
          disk (solid line) and 1000, 3000, $10\,000$, $30\,000$
          and $50\,000$ years (dotted lines from right to left) after the 
          switch off of the central radiation source.}
\label{andromeda}
\end{figure}

In Fig.~19 we show how structure in the same disk as in the previous
figures is probed by a flare of the nucleus. 
This case might be relevant for spiral galaxies like Andromeda,
where the neutral gas is very concentrated in a ring.
We map the surface brightness of the scattered emission produced
by a gas disk with a 2 kpc broad molecular 
ring at 5 kpc, with a mass surface density two times larger
than the other parts of the disk, illuminated by a flare of the central 
source of $1000$ years duration at different times after the end of
the flare. 
The velocity of the light front behind (or above) the nucleus is 
much smaller than the superluminal one in front (or below) of the source. 
As time progresses the light front scans along the molecular ring. 
GMCs will lighten up and and produce flares in the response. 
With the spatial resolution of {\it Chandra} 
one should be able to resolve these clouds in a galaxy
up to a distance of about 10 Mpc. 
The characteristic time of these events is rather short.

At a certain point the lower, faster light front reaches the 
edge of the gas disk. The ``ring'' will start to break up and 
a luminous arc travels upwards, slowly fading away as it
reaches larger distances. 

Obviously a flare due to the tidal disruption of a star
in the nucleus with a duration of the order of a few
years will produce a feature in the scattered surface brightness
distribution of smaller width than the one shown
in Fig.~19. If the tidal disruption rate
is large enough, of the order of $10^{-4}\,{\rm year}^{-1}$, 
one might find several of these arcs per galaxy.

\subsection{Scattered X-ray luminosity}
For a source too distant to be resolved by X-ray telescopes
the measurable quantity is the scattered flux from which
one can derive a scattered luminosity.
In this section we compute the scattered luminosity
of a spiral galaxy with the same neutral gas distribution
as our Galaxy in response to the above assumed
forms of activity  of its nucleus. 

\begin{figure}
\centering
\includegraphics[width=\columnwidth]{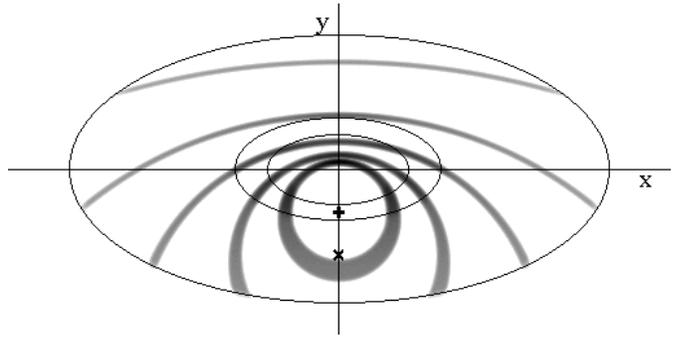}
 \caption{The surface brightness of the scattered emission produced
          by a gas disk with a 2 kpc broad molecular 
          ring at 5 kpc illuminated by a flare of the central 
          source of $1000$ years duration having occured $5000$, $10\,000$,
          $20\,000$, $40\,000$ and $60\,000$ years ago. 
          The edges of the gas disk at 16 kpc 
          and the the molecular ring
          at 4 and 6 kpc are  marked by solid ellipses.
          The inclination angle of the disk is $i=60^{\mathrm{o}}$.
          The centers of the ``rings'' $5000$ and $10\,000$ year
          after the end of the flare have been marked by a plus sign
          and a cross.
          Note how with progressing time the flare scans along
          the molecular ring. The highest surface brightness regions,
          besides the one close to the nucleus, are where the flare 
          intersects with the  molecular ring.}
\label{andromeda}
\end{figure}

Fig.~20 shows the scattered luminosity
after the AGN has switched off as a function of time
for different inclinations of the disk. 
As parameters we choose as before
$L_\mathrm{agn}=10^{44}\,{\rm erg\, s}^{-1}$
and a constant HI distribution
$\Sigma_\mathrm{HI}=4\,M_{\odot}\,{\rm pc}^{-2}$
up to 16 kpc.

The different luminosities for different inclinations
at times $t-t_\mathrm{o}\leq 0$, 
when the whole disk is illuminated,
come about because of the angular dependence 
of the differential scattering cross-section. 
For a face on disk $i=0^{\mathrm{o}}$
the scattering angle is obviously $\pi/2$.
Because the differential cross-section for Thomson scattering
is minimal for an angle of $\pi/2$ the face on disk has the
smallest scattered luminosity.
The much stronger differences at larger times
are due to the superluminal motion of the light
front. It leads to the strong effects in the low
luminosity tail.
Note the similarity of these curves with
the ones of the scattered flux produced
by the Galactic HI distribution in response to a switch off or
a flare
of Sgr A$^{*}$ (Figs.~11 and 12). The edge on case ($i=90^{\mathrm{o}}$) is 
similar to the case of the observer located 
inside the Milky Way looking at small Galactic longitudes.
The X-ray binary population of
the Milky Way produces a luminosity of about 
$2\times10^{39}\,{\rm erg}\,{\rm s}^{-1}$ in the 2-10 keV range
(Grimm et al.~2001). It will be therefore difficult to detect
the scattered signal below this limit.

\begin{figure}
\centering
\includegraphics[width=\columnwidth]{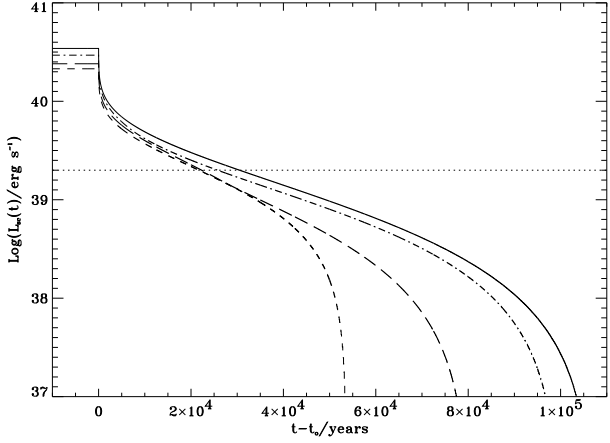}
 \caption{The scattered X-ray luminosity observable at a time t 
          produced by a spiral galaxy with the 
          HI distribution of the Milky Way
          in response to its central radiation source 
          switching off at a time $t_\mathrm{o}$. 
          The assumed luminosity of the AGN is
          $10^{44}\,{\rm erg}\,{\rm s}^{-1}$. 
          The different curves are for an 
          inclination angle of the disk of $90^\mathrm{o}$ (solid),
          $60^\mathrm{o}$(dot-dashed), $30^\mathrm{o}$ (long dashed) 
          and $0^\mathrm{o}$ (dashed). The dotted line marks
          the Galactic luminosity produced by X-ray binaries.}
\label{spirals}
\end{figure}

For the face on disk ($i=0^{\mathrm{o}}$)
one can derive a simple analytic solution for the
scattered luminosity as a function of time.
In this case the illuminated and not illuminated 
parts are concentric disks (switch off or switch on scenario) 
or annuli (flare scenario). The problem therefore is axially symmetric
and the projected distance $\rho$ is equivalent to the
physical distance $r$ from the nucleus. 
Imagine the disk to made up of small mass elements 
${\rm d}M=\Sigma(\rho)\,{\rm d}A=\Sigma(\rho)2\pi \rho{\rm d}\rho$.
If the mass element is illuminated it contributes 
a fraction ${\rm d}L\propto {\rm d}M/\rho^2$ (Eq.~14)
to the scattered luminosity. 
The total scattered luminosity at a time $t$
is just the integral over all the illuminated mass elements
\begin{equation}
L_\mathrm{sc}(t)=
\int {\rm d}L\propto
\int_{a(t)}^{b(t)}
\frac{\Sigma(\rho)\,{\rm d}\rho}{\rho}.
\end{equation}
For a constant mass surface density 
$\Sigma(\rho)=\Sigma_\mathrm{HI}$
one obtains the simple solution
\begin{equation}
L_\mathrm{sc}(t)
= 
L^\mathrm{sc}_{\mathrm{o}} \ln
\frac{b(t)}
{a(t)},
\end{equation}
with
\begin{eqnarray}
L^\mathrm{sc}_{\mathrm{o}}=
2\times 10^{39}\,{\rm erg\, s}^{-1}\,
\left(
\frac{L_\mathrm{agn}}{10^{44}\,{\rm erg\, s}^{-1}}
\right)
\left(
\frac{\Sigma_\mathrm{HI}}{4\,M_{\odot}\,{\rm pc}^{-2}}
\right).
\nonumber
\end{eqnarray}
For the face on disk
the ellipsoids travel with the velocity of light
radially outward and the the inner and outer radius for the flare scenario
can be be described as $a(t)=ct$ and 
$b(t)=c(t+\Delta)$.
For the switch off scenario one has $b(t)=\rho_\mathrm{max}$ and
for the switch on case $a(t)=0$.
This simple analytic formula helps to understand the behaviour of the
curves presented in Fig.~21, which are the results of our numerical
integration.
For large times the flare 
solutions join with the switch off solution. This happens when only 
the outer parts of the disk of size corresponding to the flare
duration are still scattering.
\begin{figure}
\centering
\includegraphics[width=\columnwidth]{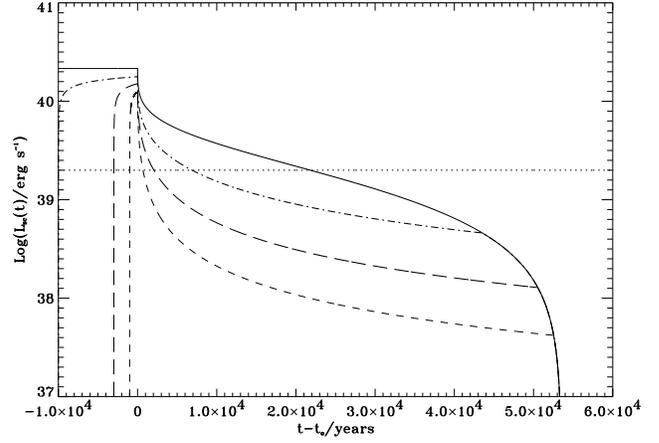}
 \caption{The scattered X-ray luminosity observable at a time t 
          produced by a face on disk ($i=0^{\mathrm{o}}$) with the 
          HI distribution of the Milky Way in response to 
          flares of the central radiation source of different
          durations ending at a time $t_\mathrm{o}$: 
          $1000$ years (dashed), $3000$ years (long dashed)
          and $10\,000$ years (dot-dashed). 
          The solid line shows the switch off case.
          The assumed luminosity of the AGN is
          $10^{44}\,{\rm erg}\,{\rm s}^{-1}$. The dotted line marks
          the Galactic luminosity produced by X-ray binaries.}
\label{flares}
\end{figure}

\subsection{Fluorescent iron K$_{\alpha}$-line luminosity}
Scattering of X-rays produced by a powerful AGN is accompanied 
by X-ray photo absorption by heavy elements and 
emission in the narrow fluorescent
K$_{\alpha}$-lines.
The iron K$_{\alpha}$-line at 6.4 keV is the brightest 
spectral emission feature in the scattered radiation. Its equivalent width
in the scattered X-ray flux is close to 1 keV for a normal cosmic
abundance of iron (Vainshtein \& Sunyaev 1980). The time dependence 
of the K$_{\alpha}$-line luminosity should repeat 
that of the scattered continuum.
The main difference is that scattering in the K$_{\alpha}$-line
is isotropic.
The results for the K$_{\alpha}$-line fluorescence line
should therefore differ by a factor of two or less and follow
similar trends as presented in the previous figures.
For example if we would plot the luminosity in the K$_{\alpha}$-line
in Fig.~20 the differences for times before the switch off would
disappear and one would get the same luminosity for all disks
independent of inclination.
Since the narrow K$_{\alpha}$-line originating in Galactic X-ray binaries
has a very low equivalent width, observations of the fluorescent,
iron K$_{\alpha}$-line are a more promising way to detect the scattered 
radiation produced by a faded flare of Sgr A$^{*}$. 

\section{Discussion}
In this paper we have presented upper limits 
on the hard X-ray 
luminosity of Sgr A$^{*}$ in the recent past.
As the scattering material we have considered molecular
and neutral atomic hydrogen present in the Galactic Disk.
The CO and X-ray data we had access to, have allowed us to 
derive limits for the 4-10 keV luminosity of Sgr A$^{*}$
down to about $8\times10^{40}\,{\rm erg\,s}^{-1}$ at certain
times. For other times the limits are of the
order of $10^{41}-10^{42}\,{\rm erg\,s}^{-1}$. 
Now we want to briefly
address a few points which demand some further discussion:

\begin{list}{}{}
\item[{\it a) Anisotropy of the X-ray emission of Sgr A$^{*}$} 
      $\,\,\,\,\,\,\,\,\,\,\,\quad$]
In simulating the activity of the GC we made the assumption 
that the emission of Sgr A$^{*}$ was isotropic. Studies of AGN show that
their radiation field is often anisotropic due to the
presence of a relativistic jet or as a result of 
shielding by a molecular torus.
A jet produced by Sgr A$^{*}$ would probably point away
from the Galactic plane. This would make our derived limits
weaker.
\item[{\it b) Chandra observations of the Galactic Ridge}
     $\,\,\,\,\,\,\,\,\,\,\,\quad$]
Ebisawa et al.~(2001) have reported the results of a deep {\it Chandra}
observation of a Galactic plane region 
($l=28.45^{\mathrm{o}}, b=-0.2^{\mathrm{o}}$),
which is devoid of known X-ray point sources.
The total X-ray flux in the 2-10 keV range is determined to be
$\sim 1.1\times 10^{-10}\,{\rm erg}\,{\rm s}^{-1}{\rm deg}^{-2}$.
About ten per cent of this flux is accounted for by point sources
resolved by {\it Chandra}. These are extragalactic sources
seen through the Galactic disk. 90 \% of the observed flux
is due to diffuse emission at the level of the sensitivity 
and the angular resolution of {\it Chandra}.
This diffuse flux is within a
factor of two of the ones measured by {\it ASCA} in 
nearby fields, which are of the order of 
$\sim 6\times 10^{-11}\,{\rm erg}\,{\rm s}^{-1}{\rm deg}^{-2}$
(Sugizaki et al.~2001).
Unfortunately none of the GMCs listed in the SRBY87
sample lies in this field,
but the HI column density in this direction is about 
$2\times10^{22}\,{\rm cm}^{-2}$, similar to the one in the
{\it ASCA} fields.
Improvements might result from the spectral data
obtained by {\it XMM-Newton}. If {\it XMM-Newton} will be 
able to detect the K$_{\alpha}$-line of neutral iron in
the same field this will open the possibility to 
separate the scattering contribution.
\item[{\it c) Chandra observations of the Orion Nebula}
     $\,\,\,\,\,\,\,\,\,\,\,\quad \quad \quad$]
The Trapezium region ($l=209.01^{\mathrm{o}}, b=-19.38^{\mathrm{o}}$)
of the Orion Nebula has recently been 
observed by {\it Chandra} (Schulz et al.~2001). An 
upper limit for the diffuse emission in the energy range
0.1-10 keV of 
$2\times10^{28}\,{\rm erg}\,{\rm s}^{-1}{\rm arc\,sec}^{-2}$
is derived, which corresponds to a diffuse flux of 
$\sim 1\times 10^{-8}\,{\rm erg}\,{\rm s}^{-1}{\rm deg}^{-2}$. 
The distance towards the Orion nebula 
is $\approx 440$ pc, corresponding to a time delay 
of $\sim2400$ years.
The galactocentric distance of Orion is $R\sim 8.8$ kpc.
The measured H$_2$ column density in this direction
is $3.2\times 10^{22}\,{\rm cm}^{-2}$,
which translates into an upper limit for the luminosity of 
Sgr A$^{*}$ of $3\times10^{41}\,{\rm erg}{\rm s}^{-1}$.
So GMCs in our vicinity might give comparable limits to the ones
derivable from GMCs located in the molecular ring, but their 
time delays will be only of the order of their distances.
\end{list}

An obvious next step would be to apply this method to constrain
the luminosity of Sgr A$^{*}$ in the past
to GMCs and HI gas with still larger distances from the GC and
thereby extending our time coverage. 
Because the scattered luminosity is proportional to
$M_\mathrm{gas}/R^{2}$, there certainly will be
a limit up to what distances we will obtain meaningful results
with this method.

A reservoir of neutral gas 
which has the potential to scatter the radiation
emitted by Sgr A$^{*}$ at even larger distances 
and therefore longer time delays is the LMC. 
With a distance of $\sim 50$ kpc from the sun
the time delay turns out to be $\sim 250\,000$ years.
Unfortunately the existing data does not permit to obtain 
interesting limits upon the X-ray luminosity of Sgr A$^{*}$. 
We therefore should wait for {\it XMM-Newton}
observations of the iron K$_{\alpha}$-line in the direction
towards massive gas complexes in the LMC.

\begin{acknowledgements}
We thank M.~Sugizaki for kindly providing data used in this work.
We are thankful for insightful letters from W.~Kegel and 
W.~Priedhorsky and acknowledge discussions with E.~Churazov, 
M.~Gilfanov and R.~Wielebinski. 
\end{acknowledgements}

\end{document}